\documentclass[11pt]{article}

\usepackage{authblk}  % have affiliations with article
\usepackage{graphicx} % Include figure files
\usepackage{epsfig}
\usepackage{dcolumn}  % Align table columns on decimal point
\usepackage{bm}       % bold math
\usepackage{amsmath}
\usepackage{amssymb}
\usepackage{url}

\usepackage{float}

\usepackage{multirow}
\usepackage{booktabs}

\newcommand{\beq}{\begin{equation}} \newcommand{\eeq}{\end{equation}}
\newcommand{\bea}{\begin{eqnarray}} \newcommand{\eea}{\end{eqnarray}}

  \newcommand
{\Romannumeral}[1]{\uppercase\expandafter{\romannumeral#1}}

\newcommand{\be}{\begin{enumerate}} \newcommand{\ee}{\end{enumerate}}
\newcommand{\bi}{\begin{itemize}} \newcommand{\ei}{\end{itemize}}
\newcommand{\ba}{\begin{array}} \newcommand{\ea}{\end{array}}
\newcommand{\bc}{\begin{center}} \newcommand{\ec}{\end{center}}
\newcommand{\bt}{\begin{tabular}} \newcommand{\et}{\end{tabular}}

\def\lsim{\mathrel{\rlap{\lower4pt\hbox{\hskip1pt$\sim$}}
    \raise1pt\hbox{$<$}}}           % less than or approx. symbol
\def\gsim{\mathrel{\rlap{\lower4pt\hbox{\hskip1pt$\sim$}}
    \raise1pt\hbox{$>$}}}           % greater than or approx. symbol

\newcommand{\half}{\textstyle {1\over2} \displaystyle}    % One half
\newcommand{\quarter}{\textstyle {1\over4} \displaystyle} % One quarter
\newcommand{\Dslash}{{\hbox{D}\kern-0.6em\raise0.15ex\hbox{/}}} % D slash

\renewcommand{\et}{\eta}

\begin{document}
\thispagestyle{empty} % suppresses display 1st page's number
	
\setlength{\oddsidemargin}{0cm}
\setlength{\baselineskip}{7mm}

\begin{normalsize}\begin{flushright}
May. 2020
\end{flushright}\end{normalsize}

\begin{center}

\vspace{15pt}

{\Large \bf Gravitational Fluctuations as an Alternative to Inflation III. 
\newline 
Numerical Results}

\vspace{30pt}

{\sl Herbert W. Hamber ${}^a$ \footnote{HHamber@uci.edu.}, 
Lu Heng Sunny Yu ${}^{a,b}$ \footnote{Lhyu1@uci.edu.},
Hasitha E. Pituwala Kankanamge ${}^{a}$ \footnote{EPituwal@uci.edu.}
} 
\\
${}^a$ Department of Physics and Astronomy \\
University of California \\
Irvine, CA 92697-4575, USA
\\
${}^b$ Theory Division \\
SLAC National Accelerator Laboratory \\
Sand Hill Road \\
Menlo Park, CA 94309, USA
\\

\vspace{10pt}
\end{center}

%  ABSTRACT 

\begin{center} 
{\bf ABSTRACT } 
\end{center}

\noindent

Power spectra play an important role in the theory of inflation, and their
ability to reproduce current observational data to high accuracy is often 
considered a triumph of inflation, largely because of a lack of credible 
alternatives.  
In previous work we introduced an alternative picture for the cosmological power spectra based on the nonperturbative features of the quantum version of Einstein's gravity, 
instead of currently popular inflation models based on scalar fields.  
The key ingredients in this new picture are the appearance of a nontrivial gravitational 
vacuum condensate (directly related to the observed cosmological constant), and a
calculable renormalization group running of Newton's $G$ on cosmological scales.
More importantly, one notes the absence of any fundamental scalar fields in this approach.  
Results obtained previously were largely based on a semi-analytical 
treatment, and thus, while generally transparent in their implementation, 
often suffered from the limitations of various approximations and simplifying 
assumptions.  
In this work, we extend and refine our previous calculations by laying out
an updated and extended analysis, which now utilizes a set of suitably modified
state-of-the-art numerical programs (ISiTGR, MGCAMB and MGCLASS) 
developed for observational cosmology.  
As a result, we are able to remove some of the approximations employed in our
previous studies, leading to a number of novel and detailed physical predictions.
These should help in potentially distinguish the vacuum condensate picture 
of quantum gravity from that of other models such as scalar field inflation.  
Here, besides the matter power spectrum $P_m(k)$, we work out in detail 
predictions for what are referred to as the TT, TE, EE, BB angular spectra, 
as well as their closely related lensing spectra.  
However, the current limited precision of observational data today (especially 
on large angular scales) does not allow us yet to clearly prove or disprove either 
set of ideas.  
Nevertheless, by exploring in more details the relationship between gravity and 
cosmological matter and radiation both analytically and numerically, together 
with an expected future influx of increasingly accurate observational data, one can 
hope that the new quantum gravitational picture can be subjected to further 
stringent tests in the near future.

% Keywords : Quantum Cosmology, Quantum Gravity, Inflationary Cosmology

% Preprints 2019, 2019100101 
% doi: 10.20944/preprints201910.0101.v1
% new doi :10.3390/universe5110216
% arXiv:1910.02990 [gr-qc]
% published in : Universe 2019, 5(11), 216; 

\newpage

%%%%%%%%%%%%%%%%%%%%%%%%%%%%%%%%%%%%%%%%%%%%%%%%%%%%%%%%%%%%%%%%%%%%
\section{Introduction}
\label{sec:intro}  

In cosmology, we know that the Universe is not perfectly homogeneous and isotropic, but rather comprises of fluctuations in matter and energy densities.  Furthermore, these fluctuations are congregated and correlated in a rather specific manner.  
Detailed measurements reveals fluctuations of various sizes follows a well-defined patterns, which can be quantified with correlation functions and power spectra
%\cite{pee93,pee98,%bah03
%bau06,%lon07,
%teg02%,teg04,dur14,wan13,coi12
%}.
[1-4].  
The question of why these density fluctuations are distributed the way they are is thus an 
important one in cosmology.
The conventional explanation for the shape of these power spectra is provided by inflation models, 
which are based on hypotheses of additional primordial scalar fields called inflatons \cite{gut81, lin82, alb82}.  
The shape of the observed power spectra are then derived from quantum fluctuation of these 
primordial inflaton fields, and the agreement of this prediction with observations 
to high accuracy has been widely regarded as a great triumph and confirmation for 
inflation \cite{lid00}. 

In our previous works \cite{hyu18,hyu19}, we have offered an alternative explanation 
based on gravitational fluctuations alone without inflation, which to our 
knowledge is the first-of-its-kind.  
While the theory of quantum gravity remains speculative in the short-distance regime -- 
due to both the infinite number of allowed higher-order operators consistent with general 
covariance together with a lack of experimental results in this regime, the long-distance or infrared limit 
of the theory is however in principle well-defined and unique, governed primarily 
by the concept of universality.  
Nevertheless, this long-distance quantum theory of gravity still suffers from 
being perturbatively nonrenormalizable, rendering perturbation theory useless for calculating any quantum corrections in gravity.
However, in the past decades, well known field theory techniques have 
been extensively developed, applied and even tested to high accuracy in various disciplines of
physics where perturbation theory fails (e.g. non-linear sigma model, Heisenberg magnets).  
It is thus highly conceivable that these nonperturbative techniques may find use 
in deriving physical consequences for another perturbatively nonrenormalizable theory such 
as gravity.  

From previous efforts \cite{ham17, book}, it was shown that quantum effects of gravity may manifest 
themselves not only on the extreme small (UV) scales, but also on the extreme large (IR), cosmological scales.  
In particular, our work \cite{hyu18,hyu19} have shown that, utilizing nonperturbative field theory methods, much of the cosmological matter power spectrum can be derived 
and reproduced purely from Einstein gravity and standard $\Lambda$CDM cosmology alone, without the need of any 
additional scalar fields as advocated by inflation.  
We have shown that not only the predictions agree quite well with recent data
by the Planck Collaboration \cite{planck18}, 
but also that additional quantum effects predict subtle deviations from the 
classical picture, which allows this approach to be testable in the near future with increasingly powerful cosmological experiments.

In this paper, we extended our analysis in two major areas.  
First, we utilized a number of current numerical cosmological programs, such as
ISiTGR, MGCAMB and MGCLASS.
Secondly, with the help of these numerical programs, we generated predictions 
for all other cosmologically significant spectra, including polarizations 
($C_l^{EE}, C_l^{BB}, C_l^{TE}$, etc.) and lensing spectra ($C_l^{\phi\phi},C_l^{T\phi}$, etc.).
The paper is organized as follows.  In Sec. \ref{sec:background}, 
we summarize the theoretical basis as relevant to present discussion.  
Sec. \ref{sec:numprog} introduces the numerical programs we use.  
Sec. \ref{sec:numres} presents the numerical results and analysis.  
Finally, key points and future work are summarized in the conclusion.

%%%%%%%%%%%%%%%%%%%%%%%%%%%%%%%%%%%%%%%%%%%%%%%%%%%%%%%%%%%%%%%%%%%%
\section{Background}
\label{sec:background}  

In this section, we will provide a brief review of the quantum theory of gravity and how it is related to various power spectra that can be measured in cosmology.  More detailed accounts of the nonperturbative approach to quantum gravity and the derivation of the spectra can be found in previous work \cite{hyu18,hyu19,ham17, book}.  The following will therefore only serve to summarize the key points and main results that are relevant for the subsequent discussion.

Quantum gravity, the covariantly quantized theory of a massless spin-two particles, 
is in principle a unique theory, as shown by Feynman some time ago \cite{fey63, fey95}, 
much like Yang-Mills theory and QED are for massless spin-one particles.  
In the covariant Feynman path integral approach, only two key ingredients are needed to 
formulate the quantum theory - the gravitational action $ S \left[ g_{\mu\nu} \right] $ 
and the functional measure over metrics $ d \left[ g_{\mu\nu} \right] $, 
leading to the generating function  
\begin{equation}
Z \left[ g_{\mu\nu} \right] =
\int  d \left[ g_{\mu\nu} \right] e^{ \frac{i}{\hbar} S \left[ g_{\mu\nu} \right] } \;\; ,
\label{eq:pathint}
\end{equation}
where all physical observables could in principle be derived from.  For gravity the action is given by the Einstein-Hilbert term appended by a cosmological constant  
\begin{equation}
S \left[ g_{\mu\nu} \right] = 
\frac{1}{16 \pi G} \int d^4 x  \sqrt{g}  \left( R - 2 \lambda \right) \;\; ,
\label{eq:action}
\end{equation}
where $R$ is the Ricci scalar, $g$ being the determinant of the metric $ g_{\mu\nu} (x) $, 
$G$ Newton's constant, and $ \lambda $ the scaled cosmological constant
(where a lower case is used here, as opposed to the more popular upper case in cosmology, 
so as not to confuse 
it with the ultraviolet-cutoff in quantum field theories that is commonly associated with 
$\Lambda$).  
The other key ingredient is the functional measure for the metric field, which in 
the case of gravity describes an integration over all four metrics,
with weighting determined by the celebrated DeWitt form \cite{dew62}. 
There are two important subtleties worth noting here.  
Firstly, in principle, additional higher derivative terms that are consistent with general 
covariance could be allowed in the action, but nevertheless will only affect physics 
at very short distances and will not be relevant nor needed here for studying large-distance cosmological effects.  
Secondly, as in most cases that the Feynman path integral can be written down, from 
non-relativistic quantum mechanics to field theories, the formal definition of integrals 
requires the introduction of a lattice, in order to properly account for the known fact 
that quantum paths are nowhere differentiable.
It is therefore a remarkable aspect that the theory, in a 
nonperturbative context, does not, at least in principle, seem to require any additional extraneous 
ingredients, besides the standard ones mentioned above, to properly define a 
quantum theory of gravity.

At the same time, gravity does present some rather difficult and fundamentally inherent 
challenges, such as its well-known perturbatively nonrenormalizable feature due to a badly 
divergent series in Newton's constant $G$, the intensive computational power 
needed for any numerical calculation due to it being a highly nonlinear theory, the conformal instability which makes the 
Euclidean path integral potentially divergent, and further genuinely 
gravitational-specific technical complications such as the fact that physical distances between 
spacetime points -- which depend on the metric which is a quantum entity -- fluctuate.  

Although these hurdles will ultimately need to be addressed in a complete 
and satisfactory way, a comprehensive account is of course far beyond the scope of this paper.  
However, regarding the perturbatively nonrenormalizable nature, some of the 
most interesting phenomena in physics often stem from non-analytic behavior in the 
coupling constant and the existence of nontrivial quantum condensates, which 
are hidden from and impossible to probe within perturbation theory alone.  
It is therefore possible that certain challenges encountered in the case of gravity are 
likely the result of inadequate perturbative treatments, and not necessarily a 
reflection of some fundamentally insurmountable problem with the theory itself.  
Here, we shall take this as a motivation to utilize the plethora of well-established 
nonperturbative methods to deal with other quantum field theories where 
perturbation theory fails, and attempt to derive sensible physical predictions 
that can hopefully be tested against observations.  
More detailed accounts on the other various issues associated with the theory of quantum 
gravity can be found for example in \cite{ham17, book}, and references therein.

For our present discussion, we will mention several main results and ingredients from this perspective.  
The nonperturbative treatments of 
quantum gravity via both Wilson's $ 2 + \epsilon $ double expansion (both in $G$ and the spacetime dimension) 
and the Regge-Wheeler lattice path integral formulation \cite{lesh84} reveal the existence of a 
new quantum phase, involving a nontrivial gravitational vacuum condensate \cite{ham17}.  
Along with this comes a nonperturbative characteristic correlation length scale, $\xi$, 
and a new set of non-trivial scaling exponents, as is common for well-studied 
perturbatively non-renormalizable theories 
$\nu$ 
%\cite{wil72,wil73,par73,par75,par85,par76}.
[18-23].  
%,par81,itz91,car96,zin02,bre10}.
Together, these two parameters characterize the quantum corrections to physical observables 
such as the long-distance behavior of invariant correlation functions, as well as the 
renormalization group (RG) running of Newton's constant $G$, which in coordinate
space leads to a covariant
$G (\Box)$ with $\Box = g^{\mu\nu} \nabla_\mu \nabla_\nu $ \cite{book}.
In particular, in can be shown \cite{ham17, cor94} that for $r<\xi$, the correlation 
functions of the Ricci scalar curvatures over large geodesic separation 
$r\equiv\left|x-y\right|$ scales as 
\begin{equation}
G_R(r) \; = \; \langle \; \delta R(x) \; \delta R(y) \; \rangle \sim \; \frac{1}{r^{2(d-1/\nu)}} \;\; ,
\label{eq:GRcorr_scaling}
\end{equation}
where $d$ here the dimension of spacetime.  
Furthermore, the RG running of Newton's constant can be expressed as
\begin{equation}
G(k) \; = \; G_0 
\left[ \, 1 + 2 \; c_0 \left( \frac{m^2}{k^2} \right)^\frac{1}{2\nu}
+ \mathcal{O} \left( \left( \frac{m^2}{k^2} \right)^\frac{1}{\nu} \right) \,
\right]
\label{eq:Grun1}
\end{equation}
where $m\equiv = 1/\xi$, as the characteristic mass scale, and $2 \, c_0 \approx 16.04$ a nonperturbative coefficient, which can be
computed from first principles using the Regge-Wheeler lattice formulation of quantum gravity
%\cite{hw05, rei10,rei11, rei14,ham15,ham93,ham00}.  
[25-31].

Here we note the important role played by the quantum parameters $\nu$ and $\xi$.  
The appearance of a gravitational condensate is viewed as analogous to the 
(equally nonperturbative) gluon and chiral condensates known to describe 
the physical vacuum of QCD, so that the genuinely nonperturbative 
scale $ \xi $ is in many ways analogous to the scaling violation parameter 
$ \Lambda_{\bar{MS}} $ of QCD.
Such a scale cannot be calculated from first principles, but should instead 
be linked with other length scales in the theory, such as the cosmological 
constant scale $\sqrt{1/\lambda}$.  
For example, note that the vacuum curvature condensate expectation value
\begin{equation}
\frac{
\langle \; \int d^4 x \, \sqrt{g} \, R \; \rangle } 
{ \langle \; \int d^4 x \, \sqrt{g} \; \rangle
}
\equiv 
\langle \, R \, \rangle 
\label{eq:R_fluc}
\end{equation}
can be related to the cosmological constant via the Einstein field equations 
\begin{equation}
\langle \,
R
\, \rangle 
=
4 \lambda \; .
\label{eq:R_vev}
\end{equation}
Thus the quantity $\xi$ can be viewed as parameterizing the gravitational vacuum condensate.  
The combination most naturally identified with $\xi$ would be 
\begin{equation}
\frac{ \lambda }{ 3} = \frac{1 }{ \xi^2} \; ,
\label{eq:xi_def}
\end{equation}
such that $ \xi \sim \sqrt{\rm{3/}\lambda} \simeq 5300 \, \rm{Mpc} $ for the observed value of $\lambda $ \cite{ham17,loops,loops2}.  
The other key quantity, the universal scaling dimension $\nu$, 
can be evaluated via a number of methods, many of which are summarized in 
%\cite{ham15,ham93,ham00,wei79,gas78,gas7802,eps,eps2,eps3,eps4,larged,htw12,reu98,reu08,lit04,lit06,reu14,fal15,fal1501,per16,gie15}.
[29-31,34-51].  
Multiple avenues point to an indication of $\nu^{-1} \simeq 3$, which here will serve as 
a good working value for this parameter; a simple geometric argument suggests 
$\nu = 1 / (d-1) $ for spacetime dimension $d \ge 4$ \cite{book}.

It should be noted that the nonperturbative scale $\xi$ should also act as an infrared (IR) 
regulator, such that, like in other quantum field theories, expressions in 
the "infrared" (i.e. as $r \rightarrow \infty$, or equivalently $k \rightarrow 0$) 
should be augmented by
\begin{equation}
\frac{1}{k^2}
\rightarrow
\frac{1}{k^2+m^2}
\label{eq:IRreg}
\end{equation}
where the quantity $m = 1/\xi \simeq 2.8 \times 10^{-4} \, h \, \rm{Mpc}^{-1}$, 
expressed in the dimensionless Hubble constant $h\simeq 0.67$ for later convenience. 
Consequently, the augmented expression for the running of Newton's constant $G$ becomes
\begin{equation}
G(k) = G_0 \left[ \; 1 + 2 \; c_0 \left( \frac{m^2}{k^2+m^2} \right)^\frac{1}{2\nu}
+ \mathcal{O} \left(\left(\frac{m^2}{k^2+m^2}\right)^\frac{1}{\nu}\right) \; \right]
\;\; .
\label{eq:Grun2}
\end{equation}
The aim here is therefore to explore areas where these predictions can be put to a test.  
The cosmological power spectra, which are closely related to correlation functions, 
and thus take effects over large distances, provide a great testing ground for these 
quantum gravity effects.

To make contact with cosmological observations, the gravitational correlation function $G_R (r)$ 
in Eq.~(\ref{eq:GRcorr_scaling}) has to be related to the cosmologically observed matter density correlation
\begin{equation}
G_\rho (r;t,t') 
\equiv 
\left\langle \, \delta_m(\mathbf{x},t) \,\, \delta_m(\mathbf{y},t') \, \right\rangle 
= \frac{1}{V} \int_{V} d^3\mathbf{z} \; \delta_m(\mathbf{x+z},t) \; \delta_m(\mathbf{y+z},t) \;\; ,
\label{eq:key}
\end{equation}
where $r=\left|\mathbf{x-y}\right|$, and
\begin{equation}
\delta_m (\mathbf{x},t) \equiv \frac{\delta\rho(\mathbf{x},t)}{\bar{\rho}(t)} 
= \frac{\rho(\mathbf{x},t) - \bar{\rho} (t) }{\bar{\rho}(t)} \; \; .
\label{eq:delta_def}
\end{equation}
is the matter density contrast, which measures the fractional overdensity, or fluctuation, of matter denstiy $\rho$ above the average background density $\bar{\rho}$.
In the literature, this correlation is more often studied in Fourier-, or wavenumber-, space, 
$G_\rho ( \mathbf{k} ;t,t') \equiv \left\langle \, \delta ( \mathbf{k} ,t) \, \delta (- \mathbf{k} ,t') \, \right\rangle$, 
via a Fourier transform. 
It is also common to bring these measurements to the same time, 
say $t_0$, so that one can compare density fluctuations of different scales 
as they are measured and appear today.  
The resultant object $P_m(k)$ is referred to as the matter power spectrum,
\begin{equation}
P_m(k) \equiv 
(2\pi)^3 \langle \; \left| \delta ( \mathbf{k} ,t_0) \right|^2 \; \rangle =
(2\pi)^3  F(t_0)^2   \langle \; \left| \Delta ( \mathbf{k} ,t_0) \right|^2 \; \rangle \;\; ,
\label{eq:key1}
\end{equation}
where $\delta( \mathbf{k} ,t) \equiv F(t) \, \Delta( \mathbf{k} ,t_0)$.  
The factor $F(t)$ then simply follows the standard GR evolution formulas as governed 
by the Freidman-Robertson-Walker (FRW) metric.  
As a result, $P_m(k)$ can be related to, and extracted from, the real-space measurements 
via the inverse transform
\begin{equation}
\begin{aligned}
G_\rho (r;t,t') 
& =
\int \frac{d^3 k}{(2\pi)^3} \; G_\rho (k;t,t') \; 
e^{-i \mathbf{k} \cdot (\mathbf{x}-\mathbf{y})} \\
& = 
\frac{1}{2 \pi^2} \frac{F(t)F(t')}{F(t_0)^2} \; 
\int_{0}^{\infty} dk \; k^2 \; P_m (k) \; \frac{\sin{(kr)}}{kr}  \;\;  .
\end{aligned}
\label{eq:FT_PktoGrhor}
\end{equation}
It is often convenient to parameterize these correlators by a so-called 
scale-invariant spectrum, 
which includes an amplitude and a scaling index, conventionally written as 
\begin{equation}
G_\rho (r;t_0,t_0 ) = \left( \frac{r_0}{r} \right) ^ \gamma \;\; .
\label{eq:gamma_def}
\end{equation}
\begin{equation}
P_m (k) = \frac{a_0}{k^s}  \;\;  ,
\label{eq:s_def}
\end{equation}
It is then straightforward to relate the scaling indices using Eq.~(\ref{eq:FT_PktoGrhor}), 
giving $s = (d-1) - \gamma = 3 - \gamma  = 1 $.  
%\begin{equation}
%s = (d-1) - \gamma = 3 - \gamma  = 1 \;\;  .
%\label{eq:stogamma}
%\end{equation} 
Note that $ G_\rho (r; t_0, t_0 ) $ is sometimes referred to as $ \xi (r) $ in the literature, 
but we will use the former to avoid confusion with the fundamental gravitational correlation length $\xi$.

To arrive at a prediction for the matter density fluctuations $G_\rho$, $P_m$ from gravitational fluctuations $G_R$, we make use of the Einstein field equations
\begin{equation}
R_{\mu\nu} \, - \, \frac{1}{2} \, g_{\mu\nu} R + \lambda g_{\mu\nu} 
\, = \, 8 \pi G \, T_{\mu\nu} \;\; .
\label{eq:EFE}
\end{equation}
In a matter dominated era, such as the one where galaxies and clusters are formed, the energy momentum tensor follows a perfect pressureless fluid to first approximation.  Hence, the trace equation reads
\begin{equation}
R - 4 \lambda = - 8 \pi G \, T  \;\; .
\label{eq:EFEtrace}
\end{equation}
(For a perfect fluid the trace gives $T = 3 p - \rho$, and thus 
$ T \simeq - \rho $ for a non-relativistic fluid.)  
Since $\lambda$ is a constant, the variations, and hence correlations, are directly related as in
\begin{equation}
\langle \; \delta R (x) \, \delta R (y) \; \rangle \, = \, 
(8 \pi G)^2 \; \langle \; \delta \rho (x) \, \delta \rho (y) \; \rangle  \;\; .
\label{eq:GRtoGrho}
\end{equation}
As described above, quantum gravity predicts that, over large distances, the scalar curvature-
fluctuations scale as $ G_R \equiv \langle \, \delta R (\mathbf{x}) \, \delta R (\mathbf{y}) \, \rangle
 \sim  {1}/{r^2} $.  This implies that the matter density fluctuations follow an analogous scaling relation
\begin{equation}
G_\rho  = \left( \frac{r_0}{r} \right)^2
\label{eq:Grhoscaling}
\end{equation}
as $r \rightarrow \infty$, within the matter dominated era, and thus $\gamma = 2$.  
From the Fourier transform in Eq.~(\ref{eq:FT_PktoGrhor}), we get
\begin{equation}
P_m (k) = \frac{a_0}{k}  \;\; 
\label{eq:Pkscaling}
\end{equation}
as $k \rightarrow 0$ in wavenumber space, in the matter dominated regime.  This result of linear scaling is a well-tested and well-supported result from decades of cosmological measurements of galaxy correlations functions \cite{gil18}.

To extend beyond the linear matter dominated regime, the trace equation alone becomes insufficient (since the trace of the energy momentum tensor for radiation vanishes), and the full tensor equation has to be used.  Furthermore, in a real universe with multiple fluid components, interactions and transient behaviors have to be taken into account, which are governed by coupled Boltzmann equations.  
However, these classical procedures are fully worked out in standard cosmology texts \cite{wei08,dod03}.  
Following \cite{wei08}, the matter power spectrum can be written in two parts -- an initial condition known as a primordial spectrum $\mathcal{R}^o_k$, and an interpolating function between the domains known as a transfer function $\mathcal{T}(k)$.  Thus, the full $P_m (k)$ beyond the galaxy domain will take the form
\begin{equation}
P_{m} (k) \, = \, C_0 \, 
\left( \mathcal{R}^o_k \right)^2 k^4 \, \left[ \mathcal{T}(\kappa) \right]^2  \;\;\;  ,
\label{eq:Pkfull}
\end{equation}
where $C_0  \equiv 4 (2 \pi )^2 \, C^2 ( \Omega_\Lambda / \Omega_M ) / 25 \, 
\Omega_M^2 H_0^4 $ is a constant of cosmological parameters, and the $k^4$ factor for convenience.  The transfer function is usually written in terms of $\kappa\equiv \sqrt{2}k/k_{eq}$, a scaled dimensionless wavenumber, with $k_{eq}$ being the wavenumber at matter-radiation-equality.  
With this decomposition, the transfer function is a fully classical  solution of the set of Friedmann and Boltzmann equations, capturing the nonlinear dynamics.  This leaves the initial primordial function, which can be parameterized as a scale-invariant spectrum 
\begin{equation}
( \mathcal{R}_k^0 )^2 = N^2 \frac{1}{k^3} \left( \frac{k}{ k_{ \mathcal{R}} } \right)^{n_s - 1}  \;\; ,
\label{eq:Rksq_param}
\end{equation}
which is only parameterized by an amplitude $N^2$ and a spectral index $n_s$.  
$ k_{ \mathcal{R} } $ is referred to as the ``pivot scale'', 
and is simply a reference scale, conventionally taken to be 
$ k_{ \mathcal{R} } = 0.05 \, \rm{Mpc}^{-1}$.  

While the transfer function $\mathcal{T} (\kappa)$ -- the solution to the highly-coupled and nonlinear set of Friedmann, Boltzmann and continuity differential equations -- is difficult to solve, it is in principle fully determined from classical dynamics.  
Moreover, assuming standard $\Lambda$CDM cosmology dynamics and evolution, a semi-analytical interpolating formula for $\mathcal{T}(\kappa)$ \cite{wei08} is known.  
As a result, if the initial spectrum $\mathcal{R}^o_k$, or more specifically the parameters $N$ and $n_s$, is set, then $P_m(k)$ is fully determined.  
To find $N$ and $n_s$, it can be done by matching.  
Since Eq.~(\ref{eq:Pkscaling}) is known to be valid in the galaxy and cluster domains, 
and Eq.~(\ref{eq:Pkfull}) is supposed to account for all wavenumber-scales, these equations should overlap in the galaxy domain.  
So by matching Eq.~(\ref{eq:Pkscaling}), which is fixed by the scaling of curvature correlation functions, with Eq.~(\ref{eq:Pkfull}) in the overlapping region, $N$ and $n_s$ can be found, thus fully normalizing $P_m(k)$.  
More precise details of this procedure, as well as detailed comparison plots with the latest observational data, can be found in our previous work \cite{hyu18}.  
The key resultant analytical prediction for $P_m (k)$ from this procedure is also reproduced here in the later plot as the solid blue curve in Fig. \ref{fig:PubPlot2_Pk}, showing almost perfect fit to all observational data for $k \gg \sqrt[3]{2 c_0} \; m \simeq 5\times 10^{-4} \, \text{Mpc}^{-1}$.

%\begin{equation}
%\mathcal{T} ( \kappa ) \simeq %\frac{\ln[1+(0.124\kappa)^2]}{(0.124\kappa)^2} 
%\left [  
%{ {1+(1.257\kappa)^2+(0.4452\kappa)^4+(0.2197\kappa)^6} 
%\over  
%{1+(1.606\kappa)^2+(0.8568\kappa)^4+(0.3927\kappa)^6} }  
%\right ]^{1/2} \; \; ,
%\label{eq:Tk}
%\end{equation}

Finally for scales of $k$ comparable to $\sqrt[3]{2 c_0} \, m$, additional quantum effects are expected to become significant, due to the nontrivial vacuum condensation nature of gravity, enough to cause deviations from the classical $\Lambda$CDM result of $P_m(k)$.  
This scale is already hinted in for example Eq. (\ref{eq:Grun1}).
These modification can again be done either analytically or relying on a program numerically.  Analytically, the effect of the RG running of Newton's constant [Eq.~(\ref{eq:Grun1})] can be included via dimensional analysis for the correct factors of $G$ to include, 
\begin{equation}
P_m (k) \,\, \rightarrow \,\, \left[ \frac{G_0}{G(k)} \right]^2 P_m (k)  \;\; ,
\label{eq:Pk_runpromote}
\end{equation}
and IR regulations rather straightforwardly as per Eq.~(\ref{eq:IRreg}), as is done in other similar quantum condensate theories such as QCD or condensed matter theories. 
More details can again be found in \cite{hyu19}.  
These results are reproduced as a plot later (Fig. \ref{fig:PubPlot2_Pk}) to compare with the fully-numerical results, showing great agreement between them.  
To obtain the latter, i.e. the numerical approach, shall form the focus and the remaining of this paper.
Following similar analysis to determining $P_m(k)$, other spectra such as the angular temperature spectrum $C_l^{TT}$, should be fully derivable from the primordial function $\mathcal{R}^o_k$, or specifically $n_s$, which is set by the scaling of gravitational curvature fluctuations $\nu$.  
In fact, many spectra are only various variations of integral transforms with different physical observable quantities, say, photon temperature and polarization, instead of mass density $\delta \rho_m$.  
A brief review of that is given in Sec. \ref{subsec:TT}.  Finally, it should be re-emphasized that in this picture, a scalar field is not an essential ingredient to determine $\mathcal{R}^o_k$.

%\begin{equation}
%C_l = 16 \pi^2 \, T_0^2 \int_{0}^{\infty}{q^2 \,dq} \;
%\left( \mathcal{R}_k^0 \right)^2
%\left[ \, 
%j_l ( qr_L ) \widetilde{F_1} (q) + j_l^\prime ( qr_L ) %\widetilde{F_2} (q) \, 
%\right]^2 \;\;  .
%\label{eq:Cl_jl}
%\end{equation} 
%\\

It should be noted that there are intrinsic uncertainties in some of the theoretical parameters as well.
The so-called ``analytical'' approach as referred to here still relies on either numerical inputs, or some analytic approximations, at earlier different stages in order to extract physical predictions such as the scaling dimensions of $G_R$ $\nu$, from Eq.~(\ref{eq:GRcorr_scaling}), or the amplitude of the first order quantum corrections for the RG running of Newton's $G$ $c_0$, from Eq.~(\ref{eq:Grun1}), from the highly nonlinear gravitational path integral of Eq.~(\ref{eq:pathint}).  
For example, from latest lattice simulations of the path integral, it is found $1/ \nu \simeq 2.997(9) $ and $c_0 \simeq 8.02$, with the latter an error that is estimated at around $50 \%$.  
Other methods, summarized in \cite{ham17}, including observational data as studied in \cite{hyu19}, all support the value $\nu \sim 1/3$.  
As eluded in this paper as well, this is not surprising given the universality nature of this index $\nu$.  
On the other hand, the amplitude for quantum corrections $c_0$, while should remain some order 1 parameter, cannot be claimed to the same degree of confidence as $\nu$.  
For example, in the comparison with latest observational data for $P_m (k)$ in \cite{hyu19}, it seems to best fit a value roughly 7 times smaller ($c_0 \approx 1.146$).  
However, it should also be pointed out the theoretical expression defining $c_0$ possesses a slight degeneracy with the correlation length scale $\xi$ (Eq.~(\ref{eq:Grun1})).  
Hence, the data can also be interpreted as suggested a value of $\xi \sim 14000$ Mpc, around $2.5$ time larger than the expected $\xi \sim \sqrt{3/\lambda}$, or, some combination of both instead.  
In principle, the inclusion of IR regulation to the final expressions (Eq.~(\ref{eq:Grun2})) changes the shape of the curve and can in principle break the degeneracy, the current crudeness of the observational data in those regimes of $k$ is much too uncertain to make any conclusions as to the more favorable possibility.  
While we will continue to primarily refer to studying the constraints on $c_0$ for simplicity for the rest of this paper, it should be kept in mind the possibility of this degeneracy.  
It is also hopeful that with increasingly precise observational data in the future, complimented with looking at independent and orthogonal observables that we are to present in this paper, a better constraint on these theoretical parameters can be found.

Finally, it should be noted that the current most popular approach to explain the shape, or more precisely, the index $n_s$ of the matter power spectrum is typically reliant on the fluctuations of postulated primordial scalar fields from inflation models \cite{ste04}.  Given the long interest for understanding this spectral index \cite{hz70,hz72,pee70}, the ability to derive this index, as well as the lack of competing theories, is thus championed as a triumph of inflation.  The picture reviewed here, where the correlations are explained by nonperturbative critical scaling behaviors of gravitational fluctuations, is thus first-of-its-kind.  As discussed in this background, the formulation of this picture is in principle rather constricted with little flexibility.  As a result, this gravitational picture makes concrete predictions that can be concretely tested (or falsified), without suffering from the typical flexibilities in scalar-field-driven inflation models, and thus offering a compelling alternative to the canonical inflation picture.

Having reviewed this analytic background, we will next present the numerical programs we used, and the subsequent results for the cosmological spectra from effects of quantum gravity.

\vspace{20pt}

%%%%%%%%%%%%%%%%%%%%%%%%%%%%%%%%%%%%%%%%%%%%%%%%%%%%%%%%%%%%%%%%%%%%
\section{Numerical Programs}
\label{sec:numprog}  

There are a variety of publicly available Boltzmann-Einstein (EB) solvers that have been in use for the past two decades starting with CMBFAST \cite{code18}. The main independent programs are CAMB \cite{lew99} and CLASS \cite{les11} which solve the coupled Einstein-Boltzmann equations in a background FRW metric. These codes are computed for $\Lambda$CDM cosmology with a limited set of choices for a   parameteriazation of equation of state for the Dark Energy ($w$). In all our programs we use $w=-1$ ,which considers dark energy as a vacuum energy. 

\begin{figure}
\begin{center}
\includegraphics[width=1.00\textwidth]{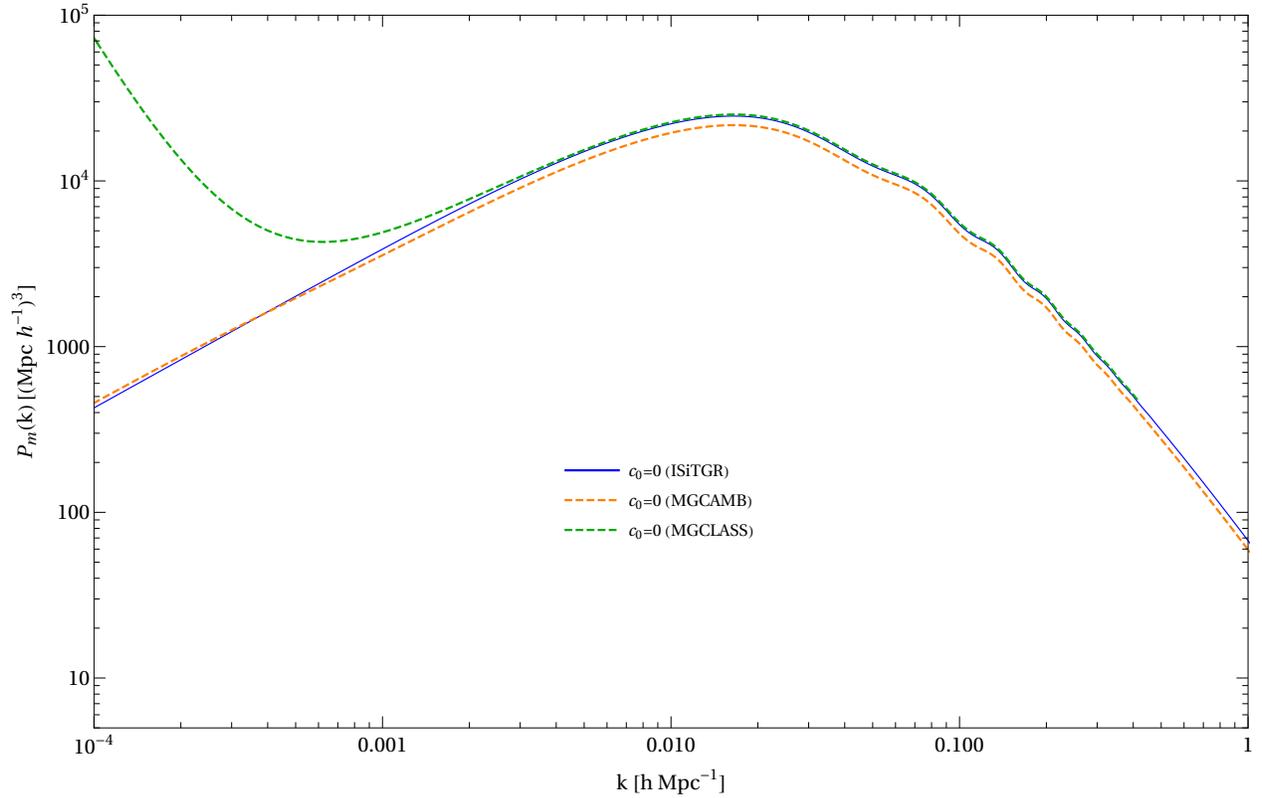}
\end{center}
\caption{
As an example we illustrate the $P_m(k)$ predictions between the three programs -- ISiTGR (blue), MGCAMB (orange), and MGCLASS (green) -- with their corresponding patches for a modified Newton's constant.  This serves as a consistency check between the programs and a validity check for their patches.  The solid curves are generated from the respective original $\Lambda$CDM programs, while the dashed curves are generated by each program's modified Newton's constant patch setting $\mu(a,k) \equiv G_\text{mod}/G_{Newt} = 1$.  It can be seen that ISiTGR is the most consistent, and hence reliable program of the three, to investigate the effects of a modified Newton's constant.
}
\label{fig:PubPlot1_P(m)compare}
\end{figure}

\begin{figure}
\begin{center}
\includegraphics[width=1.00\textwidth]{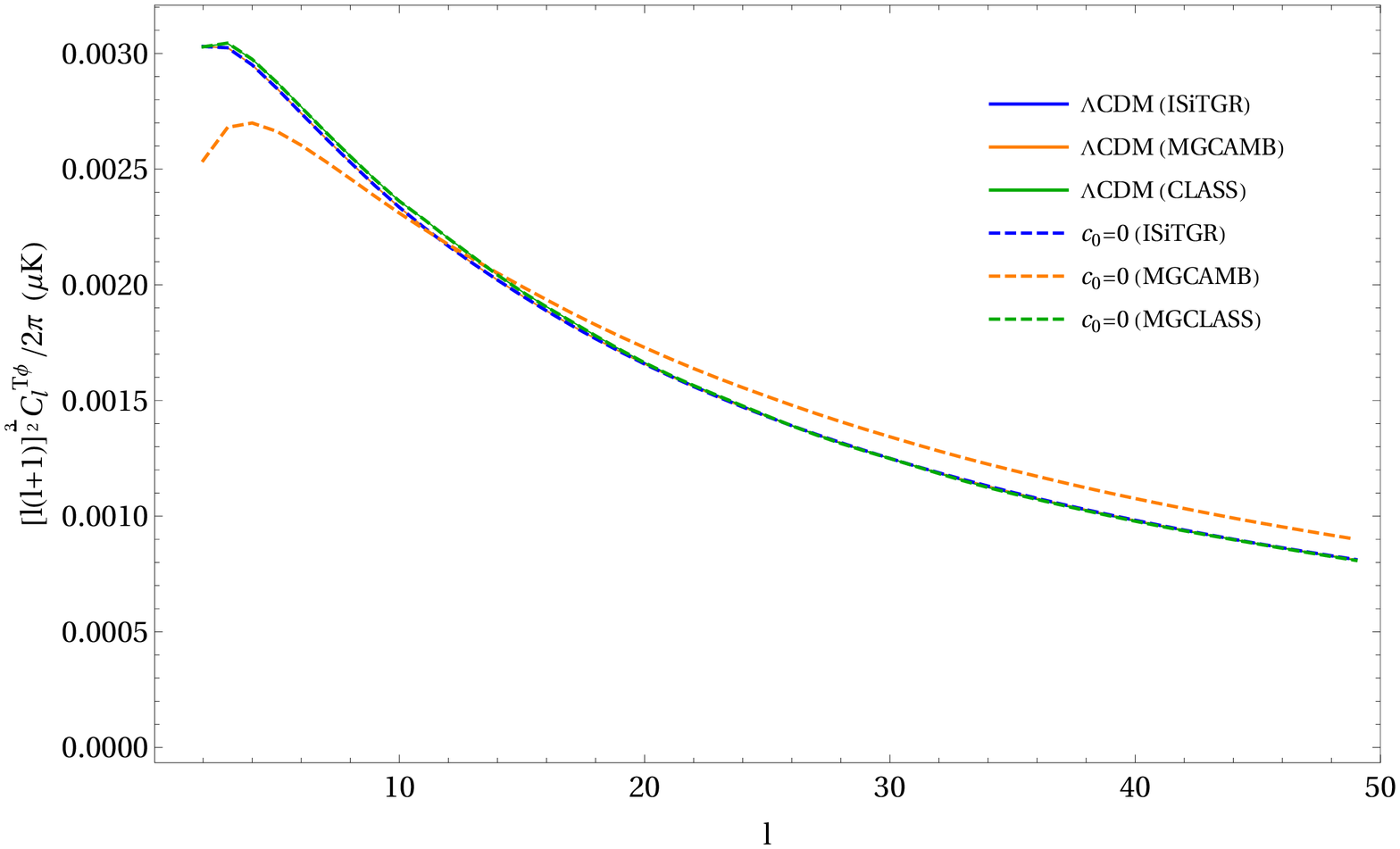}
\end{center}
\caption{
Comparison of the $C_l^{T\phi}$ predictions between the three programs -- ISiTGR (blue), MGCAMB (orange), and MGCLASS (green) -- with their corresponding patches for a modified Newton's constant.  This serves as a consistency check between the programs and a validity check for their patches.  The solid curves are generated from the respective original $\Lambda$CDM programs, while the dashed curves are generated by each program's modified Newton's constant patch setting $\mu(a,k) \equiv G_\text{mod}/G_{Newt} = 1$.  It can be seen that ISiTGR is the most consistent, and hence reliable program of the three, to investigate the effects of a modified Newton's constant.
}
\label{fig:PubPlot1_TPhi_compare}
\end{figure}

For modifications of gravity with a scale dependent gravitational constant, there are three EB solvers. We used ISiTGR (Integrated Software in Testing General Relativity) \cite{gar20} as the primary code to generate power spectra. Then we compare with another two programs MGCAMB (Modified Growth with CAMB) \cite{zucca2019mgcamb} and MGCLASS (CLASS version for phenomenological modified gravity) \cite{Baker_2015}. ISiTGR  and MGCAMB are patches for CAMB and COSMOMC \cite{lew02} which was written in the FORTRAN language, while MGCLASS  is a patch for CLASS written in C. All three programs have implemented the parameterization effective gravitational coupling $(\mu)$ - gravitation slip parameter $(\eta)$ which sometimes is denoted as $\gamma$. Those two parameters are defined as 
$\mu(a,k) \equiv G(a,k)/G_0 $ and $\eta(a,k) \equiv \Phi/\Psi  $ where $G_0$ is the laboratory value of Newton's gravitational constant and $\Phi,\Psi$ are scalar potentials in the conformal Newtonian gauge. The comparison of the three programs for no RG running of $G$ as in standard $\Lambda$CDM cosmology is shown in Fig. \ref{fig:PubPlot1_P(m)compare} and Fig. \ref{fig:PubPlot1_TPhi_compare}.

% Fig. 1

One can see that while all three program's $\Lambda$CDM prediction are generally consistent, only ISiTGR's modified Newton's constant patch with $\mu(a,k) \equiv G_\text{mod}/G_{Newt} = 1$ [or equivalently $c_0 = 0$ in Eq.~(\ref{eq:Grun2})] is consistent with its original default-$\Lambda$CDM prediction. Matter power spectrum from MGCLASS has a noticeable upper trend for small k from the $\Lambda$CDM curve, as shown in the left plot in Fig. \ref{fig:PubPlot1_P(m)compare}.  Fig.\ref{fig:PubPlot1_TPhi_compare}
 shows a significant deviation of MGCAMB's $C_l^{T\phi}$ from the $\Lambda$CDM curve. Primarily due to this reason we chose ISiTGR over these two other programs.  

% Fig. 2

%- Comment: The plot for $C_l^{\phi\phi}$ is chosen particularly because... (it most illustrates the deviations... and hence confidence for ISiTGR...)

%- Comment: The plots are generated with latest cosmological parameters as published by Planck 2018 [cite:XXXplanck18paper].

In the ISiTGR program all times are in conformal time, as is the case for CAMB. The growth equations are written based on a perturbed FLRW metric in the Newtonian gauge,
\begin{equation}
ds^2
= a(\tau)^2 \left[-(1+2\Psi)d\tau^2 + (1-2\Phi)\gamma_{ij}dx^idx^j  \right], 
\;\; 
\label{eq:metric}
\end{equation}
Where $\Phi$ and $\Psi$ are scalar gravitational potentials , $x_i$ represents comoving coordinates and $a(\tau)$ is scale factor at conformal time $\tau$. For a flat universe the three dimensional spatial metric $\gamma_{ij}$ in cartesian coordinates is given by
\begin{equation}
\gamma_{ij}
= \delta_{ij}, 
\;\; 
\label{eq:gamma metric}
\end{equation}
From now on we only discuss cosmology for a spatially flat universe, to which $k=0$ .

There are four built in functional forms for selected modified cosmologies \cite{gar19} and we used  $(\mu)$- gravitation slip parameter $(\eta)$ form. The modified growth equations are
\begin{equation}
k^2\Psi
=
-4 \pi \, Ga^2\mu(a,k)\sum_{i} \left[\rho_i\Delta_i + 
3\rho_i\left(1+w_i \right)\sigma_i  \right] ,
\;\; 
\label{eq:growth eq1}
\end{equation}
and
\begin{equation}
k^2\left[\Phi-\eta(a,k)\Psi \right]
=
12 \pi \, Ga^2\mu(a,k)\sum_{i} 
3\rho_i\left(1+w_i \right)\sigma_i  
\;\; 
\label{eq:growth eq2}
\end{equation}
where $w_i$ and $\rho_i$ are respectively the equation of state and density of $i^{th} $ particle species. Generally there are three species which are radiation, non relativistic matter and dark energy. And $\Delta_i$ is the gauge-invarient, rest-frame overdensity defined by,
\begin{equation}
\Delta_i 
= \delta_i + 3H\frac{q_i}{k} \;\; ,
\label{eq:overdensity}
\end{equation}
where $H = \dot{a}/a$ is the Hubble's constant in conformal time, fractional overdensity $\delta_i = \delta \rho/\bar{\rho} $ and $q_i$ is the heat flux, related with the peculiar velocity ($\theta_i$) 
\begin{equation}
q_i 
= \theta_i \frac{1+w_i}{k} \;\; .
\label{eq:qi}
\end{equation}
From the conservation of energy-momentum tensor of the perturbed matter fluids and for uncoupled fluid species $\Delta_i$ evolution is given by
% reference 60 in isitgr overview paper
\begin{equation}
\Delta_i
= 3(1+w_i) \, \left( \dot{\Phi} + H\Psi \right) + 3Hw_i\Delta_i - \left[ k^2+3(H^2-\dot{H})\right] \,\frac{q_i}{k} 
- 3H(1+w_i)\sigma_i
\;\;  .
\label{eq:Delta evolution}
\end{equation}
Secondary effects considered by ISiTGR are  reionization,weak gravitational lensing and the ISW effect. For reionization it uses same approach as in  CAMB \cite{lewiscamb}, namely a simple tanh model for reionization fraction $(x_e)$, given by
%tanh formula with optical depth
\begin{equation}
x_e(y) =  \frac{f}{2} \left[ 1 + \tanh{\frac{y(z_{re})-y}{\Delta_y}} \right]
 \;\; ,
\label{eq:re-ionization tanh}
\end{equation}
where $y(z)=(1+z)^{3/2}$ , $z_{re}$ is the red shift value where the $x_e = f/2 $, and $\Delta_y$ is the fractional change in y. 
The latter agrees with a Thompson scattering optical depth for an instantaneous reionzation which occurred at $z_{re}$. 
The treatment of weak lensing is discussed here later in section 4.

% Table for parameter values
\begin{table}

\caption{Values used here for cosmological parameters in the $\Lambda$CDM model. We have used the Planck-18 68\% interval from CMB power spectra, in combination with CMB lensing reconstruction and Baryonic Acoustic Oscillations (BAO).  }

\begin{tabular}{  m{7.5cm}  m{2.0cm}  m{4.5cm}  } 
\toprule
Parameter & Symbol & Value   \\ 
\hline
barryon density & $\Omega_bh^2$ & $ 2.242 \times10^{-2}$  \\ 

cold dark matter density & $\Omega_ch^2$ & $1.1933 \times10^{-1}$   \\ 

acoustic scale angle & 100 $\theta_*$ & 1.04  \\  

scalar amplitude & $A_s$ & $2.105\times10^{-9}$  \\  

reionization optical depth & $\tau$ & $5.61\times10^{-2}$  \\ 

scalar tilt & $n_s$ & $0.9665$  \\ 
\hline  
Hubble constant & $H_0$ &  67.66 $km s^{-1}Mpc^{-1}$ \\

curvature density & $\Omega_k$ & 0 \\

effective extra relativistic degrees of freedom & $N_{eff}$ & 3.046 \\

CMB temperature & $temp\char`_cmb$ & 2.7255 K \\

equation of state of dark energy & w & -1 \\ 

\bottomrule
\end{tabular}

\label{table:1}
\end{table} 
%%%%%%%%%%%%%%%%%%%%%%%%%%%%%%%%%%%%%%%%%%%%%%%%%%%%%%%%%%%%%%%%%%%%

Since the required formulation for $\mu(a,k)$ does not appear as an inbuilt function, we added a part with newly defined functions $\mu(a,k)$, $\dot{\mu}(a,k)$ for our need in the above equations. 
In accordance with Eq.~(\ref{eq:Grun2}) we have
\begin{equation}
\mu(a,k) =   1 + 2  c_0 \left( \frac{m^2}{k^2+m^2} \right)^\frac{1}{2\nu}
 \;\; ,
\label{eq:mu}
\end{equation}
and $\dot{\mu}(a,k) = 0$.

As secondary effects ISiTGR considers reionization,weak gravitational lensing and the ISW effect.$\eta(a,k)= 1$ is assumed since there is no different modifications to the potentials. ISiTGR has two binning methods but here we only used the traditional binning method. For all the power spectra computations we set the tensor part to zero. The program computes 2-point self- and cross-correlation functions for the temperature, E-mode and B-mode polarization and weak lensing potential. Each generated power spectrum appears in two separate files, one with lensing and the other without. In the following we use power spectra with gravitational lensing included.  
The values of the cosmological parameters we used here as initial conditions are shown in Table \ref{table:1}. 

In a previous paper \cite{hyu18,hyu19} we used semi-analytic methods to solve for the matter power spectra using semi-numerical approximations for the relevant transfer functions. But in the current approach the numerical programs solve the full set of Boltzmann equations, and uses integration techniques such as adaptive Runge-Kutta method to integrate all the tightly coupled equations. Secondary effects accounted for like reionization and integrated Sachs-Wolfe (ISW) effect are treated as a more general case compared to our previous work. 
%check 

\vspace{20pt}

\section{Numerical Results}
\label{sec:numres}  

In this section we present numerical results for the quantum gravitational corrections to the various cosmological spectra ($P_m (k), \, C_l^{TT}, \, C_l^{TE}, \, C_l^{EE}, ...$).  This includes both the effects of an RG running of Newton's constant and the IR regulation, obtained by replacing
\begin{equation}
G \, \rightarrow \,\, G(k) =  G_0 \left [ \;
1 + 2 \, c_0 \left( \frac{m^2}{k^2+m^2} \right)^{3/2} 
+ \mathcal{O} \left( \left( \frac{m^2}{k^2+m^2} \right)^{3} \right)
\right]  \;\;  .
\label{eq:promoteGrun}
\end{equation}
For simplicity, of the three numerical programs used for our analysis (ISiTGR, MGCAMB, MGCLASS), only the results from the ISiTGR numerical program shall be plotted.  
The reason for this choice is that we expect this program to provide better consistency and reliability in the particular region considered (small $k$, small $l$), as explained in Sec. \ref{sec:numprog}.  
Furthermore, all numerical results presented here are generated using the latest values of the cosmological parameters as given by Planck (2018) \cite{planck18}. 

In the above, $c_0$ is the coefficient that governs the amplitude of quantum corrections.  For all the following spectra, three different values of $c_0 = 0, \, 1.146$ and $8.02$ will be plotted.  Lattice calculations give $c_0 \approx 8$.  However, being a non-universal parameter, it can depend on specific choices arising from the way an ultraviolet cutoff is imposed.  
Therefore, not too much weight should not be placed on this specific value, beyond perhaps the order of magnitude.  In practice, this value could be further constrained by experiments, which is precisely what these observations of cosmological spectra can achieve.  From previous work \cite{hyu19}, using the approximate semi-analytical methods, we see that a value of $8.02/7 \simeq 1.146$ is generally favored.

%%[[Pk]]
\subsection{Matter Power Spectrum $P_m(k)$}
\label{subsec:Pk}  

\begin{figure}
\begin{center}
\includegraphics[width=0.90\textwidth]{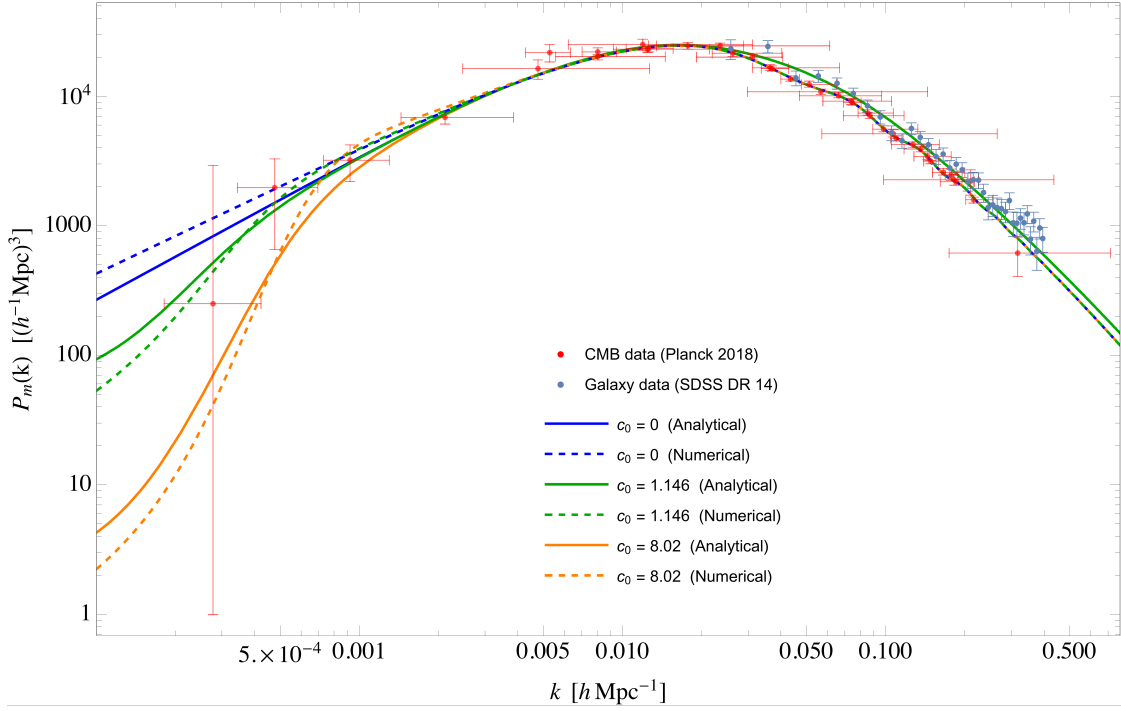}
\end{center}
\caption{
Comparison between the analytical vs. numerical predictions of the RG running of Newton's constant's effect 
on the matter power spectrum $P_m(k)$.  The solid curves represent the analytical predictions, with the 
top (blue), middle (green), and bottom (orange) representing the quantum amplitude quantum amplitudes [see Eq.~(\ref{eq:Grun2})] $c_0 = 0, 1.146, 8.02$ respectively [see Eq.~(\ref{eq:Grun2})], of which 
their detailed derivations can be found in \cite{hyu18,hyu19}.  The corresponding dashed curves 
represent the corresponding the numerical predictions generated by ISiTGR, showing very good 
general consistency with the trend derived from analytical methods.  
The observational CMB and galaxy data points, taken from the Planck (2018) collaboration 
\cite{planck18} and Sloan Digital Sky Survey (SDSS)'s 14th Data Release (DR14) \cite{gil18} are also shown.
}
\label{fig:PubPlot2_Pk}
\end{figure}

We start with the matter power spectrum $P_m(k)$.  Recall the definitions
\begin{equation}
G_\rho (r;t,t') 
\equiv 
\left\langle \, \delta_\rho(\mathbf{x},t) \,\, \delta_\rho(\mathbf{y},t') \, \right\rangle 
\; , \;\;\;
P_m (k) \sim \langle \; \delta_\rho ( \mathbf{k} ) \; \delta_\rho ( -\mathbf{k} ) \; \rangle \;\;\;,
\label{eq:Pk_def2}
\end{equation}
where the variable 
$\delta_\rho \equiv { \left( \rho - \bar{\rho} \right) }/{\bar{\rho}} $ is the fractional density fluctuations above the average, referred to in cosmology as the mass-density contrast.  
The numerical results for $P_m(k)$ obtained from the numerical program (ISiTGR), for both the classical $\Lambda$CDM (i.e. $c_0 = 0$) and quantum ($c_0 > 0$) results, as well as the respective analytical results (as derived in \cite{hyu18,hyu19}), are shown and compared in Fig. \ref{fig:PubPlot2_Pk}.

% Fig. 2

From Fig. \ref{fig:PubPlot2_Pk}, we see that all the numerical results are generally consistent with the corresponding analytical results from earlier work, which were obtained by following the semi-analytical interpolating formulas from \cite{wei08}, and the implementation of the RG running following dimensional arguments.  The small deviations may be attributed to the slightly older values of cosmological parameters  \cite{wmap03} and some analytic approximations used by Weinberg and Dicus' interpolating formula for the transfer function in \cite{wei08}, whereas the numerical results presented here use the latest cosmological parameter values from the Planck collaboration \cite{planck18}.  Despite the small discrepancies, we see that the overall trends, and the extra downwards bend due to the inclusion of the (IR regulated) RG running of Newton's constant, as predicted analytically using the semi-analytical formulas are in very good agreement with the numerical predictions using the latest fitted cosmological parameters.  This overall general agreement between the analytical and numerical result provides a good verification and confidence that the procedure of including a running Newton's constant as presented above is reliable.

The same numerical analysis has now been repeated with the other two numerical programs MGCAMB and MGCLASS, besides ISiTGR.  The result of MGCAMB is in extremely good agreement with ISiTGR, with its predictions for all three values of $c_0$ almost completely overlapping with ISiTGR's result, giving additional confidence to the latter.  However, while MGCLASS is relatively consistent with ISiTGR for most of the angular spectrum results (as we will discuss later), its prediction for $P_m (k)$ shows a rather radical upturn below $k=10^{-3}$, which is at odds with both ISiTGR and MGCAMB, as well as the analytical predictions (also shown and discussed earlier in Fig. \ref{fig:PubPlot1_P(m)compare} and then in Sec. \ref{sec:numprog}), even for the $c_0=0$ $\Lambda$CDM case.  The pathological upturn at small-$k$ and resultant disagreement of MGCLASS (even with CLASS, the original $\Lambda$CDM program that MGCLASS is based on, when setting $c_0=0$) suggests some potentially unresolved issues in MGCLASS's prediction for $P_m (k)$, while the consistent results between ISiTGR, MGCAMB, and the analytical predictions should be treated in our opinion with a higher reliability.

Given the more confident, and in principle more accurate, predictions from the numerical programs as shown in Fig. \ref{fig:PubPlot2_Pk}, it can be seen that the value of $c_0 \simeq 1.146$ is a better overall fit to the observational data from Planck, which is a consistent conclusion from our previous work that was based exclusively on the early analytical results.  Armed with the new tools of numerical programs, we will now move on to present the numerical results for the other various correlation functions, which will hopefully shed new insights to the validity of the quantum gravity effects in cosmology. 

It should be noted that there is a slight degeneracy between $c_0$ and $\xi$ in the original expression for the RG running of Newton's $G$, as in Eq.~(\ref{eq:Grun2}).  A support of a smaller $c_0$ from the observational data can equivalently be mimicked by an increase in the vacuum condensate scale $\xi$.  In fact, the apparently better fit value of $c_0 \simeq 1.146$, seven times smaller than the lattice predicted value of $\approx 8.0 \pm 3.1$, can be mimicked by simply a factor of $\sim 1.9$ larger in $\xi$.  Technically, including IR regulation will change the shape of the curve and break the degeneracy, which in principle could be fitted sophisticated say with a Monte Carlo simulation.  However, not only is that currently beyond the scope of this paper, the lack of and crudeness of data points in the small-$k$ regime will not render the exercise fruitful.

On the other hand, it may be instructive, amongst other physical motivations, to look at the quantum effects on a variety of other spectra of cosmological significance with these numerical programs.  With independent quantities and measurements, the new plots may either provide additional constraints to these quantum gravitational parameters, but also potential insights to the physics.

\subsection{Angular Temperature Power Spectrum $C_l^{TT}$}
\label{subsec:TT}

\begin{figure}
\begin{center}
\includegraphics[width=0.90\textwidth]{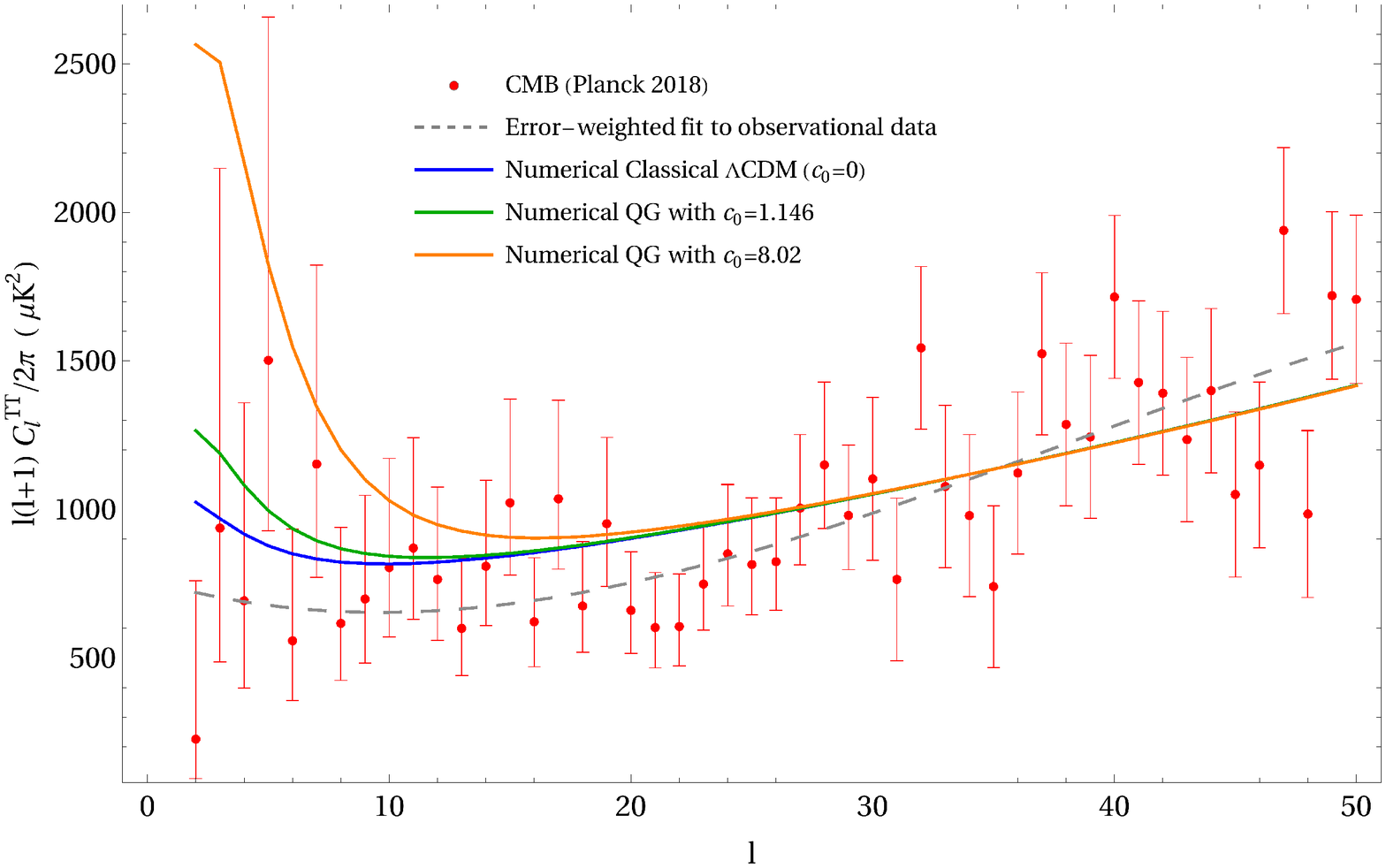}
\end{center}
\caption{
Comparison of the numerical prediction of the classical $\Lambda$CDM program vs. the numerical 
predictions 
of the RG running of Newton's constant's effect on the temperature (TT) power 
spectrum 
$C_{l}^{TT}$.  The solid curves represent the numerical predictions generated by 
the ISiTGR 
program, with 
the bottom (blue), middle (green), and top (orange) representing quantum amplitudes 
[see Eq.~(\ref{eq:Grun2})] $c_0 = 0, 1.146, 8.02$ respectively [see Eq.~(\ref{eq:Grun2})], 
showing a higher trend at large angular scales ($l<20$) as compared to the classical 
$\Lambda$CDM (no running)'s 
numerical curve.  
The dashed curve represents an error-weighted cubic fit to the observational CMB data, from 
the Planck (2018) collaboration \cite{planck18}.
}
\label{fig:PubPlot3_ClTT}
\end{figure}

The TT power spectrum is one of the most important cosmological spectrum since it is measured to high degree of accuracy, thus allowing for great insights in constraining various cosmological models.  
Fig. \ref{fig:PubPlot3_ClTT} shows the numerical predictions for the temperature-temperature (TT) angular power spectrum $C_l^{TT}$ with and without the quantum effects.  
We will first briefly recall the definitions for $C_l^{TT}$ and how  theoretical predictions for it can be obtained, and then compare them against observational data.  
Following notations in Weinberg \cite{wei08}, the temperature fluctuations $\Delta T$ can first be resolved into spherical harmonics $Y_l^{m} (\hat{n})$'s,
\begin{equation}
\Delta T (\hat{n}) 
\, \equiv \,
T (\hat{n}) - T_0
\, = \;
\sum_{lm} \, a^T_{lm} \, Y_l^{m} (\hat{n}) 
\;\; ,
\label{eq:alm_def}
\end{equation}
where $T(\hat{n})$ is the temperature in the direction $\hat{n}$, $T_0 \equiv (1/4\pi) \int d^2 \hat{n} \,\, T(\hat{n})$ the average temperature over the sky, and the coefficients $ a^T_{lm}$ quantifying the fluctuation for each harmonic.  Since $\Delta T$'s are real, and the products of $\Delta T$'s are rotationally invariant, one has
\begin{equation}
\left\langle \, \Delta T (\hat{n}) \, \Delta T (\hat{n'}) \, \right\rangle
\, = \,
\sum_{lm}  C_l^{TT} \, Y_l^{m} (\hat{n}) Y_l^{-m} (\hat{n'})
\, = \,
\sum_{l} \, C_l^{TT}  \left( 
\frac{2l+1}{4 \pi}
\right)
L_l (\hat{n} \cdot \hat{n'})
\;\;\; .
\label{eq:TT_corr}
\end{equation}
Here the $L_l$ are the Legendre polynomials, and $C_l^{TT}$ is defined as
\begin{equation}
\left\langle \,  a^T_{lm} \, a^T_{l'm'} \, \right\rangle
\equiv
\delta_{ll'} \delta_{m {-m'}} \, C_l^{TT}  \; ,
\label{eq:ClTT_def}
\end{equation}
the 2-point correlation functions of $ a^T_{lm}$, the temperature fluctuation in ``$l$"-space.  Or equivalently,
\begin{equation}
C_l^{TT}
=
\frac{1}{4 \pi}
\int d^2 \hat{n} \, d^2 \hat{n'} \;
\left\langle \, \Delta T (\hat{n}) \, \Delta T (\hat{n'}) \, \right\rangle
L_l (\hat{n} \cdot \hat{n'})
\;\; ,
\label{eq:ClTT_invert}
\end{equation}
by inverting the transformation.
As a result, the correlations for temperature-temperature fluctuations are fully quantified with the $C_l^{TT}$'s.
(Note that here we use $L_l$ instead of the usual notation $P_l$ for the Legendre polynomials, in order to avoid confusion with the matter power spectra.)

Theoretically, since CMB photon temperatures and matter density are coupled, the $C_l^{TT}$'s are therefore related to the matter power spectrum $P_m (k)$, via integral transforms that involve spherical Bessel functions and appropriate form factors and transfer functions.  However, from transforming the predictions from one set of observable to another, new insights, and potential constraints to the theory, can be derived.  

To do so, one can first use the Friedmann and continuity equations to relate the temperature fluctuations to the metric perturbations, via suitable form factors $ F_{1,2}(q) $, through
\begin{equation}
\left( 
\frac{\Delta T ( \hat{n} )}{T_0} 
\right)
= 
\int d^3 q \; e^{ i \mathbf{q} \cdot \hat{n} \, r(t_L) }
\left[ \, 	F_1(q) + i \hat{q} \cdot \hat{n} \; F_2 (q)\, \right] \;\; ,
\label{eq:DeltaT_planew}
\end{equation}
where the latter are defined as 
\begin{equation}
F_1 (q) = 
- \frac{1}{2} \, a^2 (t_L) \ddot{B}_q (t_L) - \frac{1}{2} 
a (t_L) \, \dot{a} (t_L) \dot{B}_q (t_L) 
+ \frac{1}{2} E_q (t_L) + \frac{\delta T_q (t_L)}{\bar{T} (t_L)}  \;\; ,
\label{eq:F1_general}
\end{equation}
\begin{equation}
F_2 (q) = - q \left(
\frac{1}{2} \, a (t_L) \dot{B}_q (t_L) + 
\frac{\delta u_{\gamma q} (t_L)}{ a (t_L)} \right) \;\; .
\label{eq:F2_general}
\end{equation}
Here the $ B $ and $ E $ functions are suitable decompositions of the metric perturbations, 
and $ \delta u_\gamma $ is the velocity potential for the CMB photons.  
It is known that these form factors simplify in certain gauge choices.  
In the synchronous gauge, one has $ E=0 $, whereas in the Newtonian gauge 
$ B=0 $ and $ E = 2 \Phi $, which then gives
\begin{equation}
F_1 (q) = \Phi_q (t_L) + \frac{\delta T_q (t_L)}{\bar{T} (t_L)} \;\;  ,
\label{eq:F1_Newt}
\end{equation}
\begin{equation}
F_2 (q) = -  \frac{\delta u_{\gamma q} (t_L)}{ a (t_L)}  \;\; .
\label{eq:F2_Newt}
\end{equation}
(Note that $ F_1 (q) $ and $ F_2 (q) $ are referred to as ``$ F(q) $'' and ``$ G(q) $'' 
respectively in \cite{wei08}.
Here we will use the former in order to avoid confusion with the expression for the 
running of Newton's constant $ G(k) $, as it will be implemented below.  
The above equations also assumed a sudden transition to opacity on the CMB at a time $t_L$, which nevertheless does not change the form of the basic equations and only some of the details, which are later taken into account fully with the numerical programs, discussed below.)

Hence, given appropriate initial conditions, the functions $ \Phi $ and $ \delta u_{ \gamma } $, as well as the scale factor $ a(t) $ and the function $ T(t) $, can all be obtained as solutions of the classical Friedmann equations.  These are then combined with the Boltzmann transport equations, as is done in standard cosmology, which eventually leads to unambiguous predictions for the $ C_l $'s.  
The solutions for $ F_{1,2}(q) $ can be parameterized in terms of transfer functions 
$ \mathcal{T}(\kappa) $, $ \mathcal{S}(\kappa) $ and $ \Delta(\kappa) $, leading to the following expressions for $F_1 (q)$ and $F_2 (q)$ 
\begin{equation}
F_1 (q) = 
\frac{ \mathcal{R}_q^o }{ 5 } 
\left[  
3 \, \mathcal{T} \left( 
\frac{q \, d_T }{a_L} \right) R_L \,\, - \,\,  (1 + R_L)^{ -\frac{1}{4} } 
\; e^{ - \left( \frac{q \, d_D }{ a_L } \right)^2 } 
\mathcal{S} \left( \frac{q \, d_T }{a_L} \right)
\; \cos \left[ { \frac{q \, d_H }{ a_L } + \Delta \left( \frac{q \, d_T}{a_L} \right) }  \right]
\right]   \;\; ,
\label{eq:F1_sol}
\end{equation}
\begin{equation}
F_2 (q) = \sqrt{2} \; \frac{ \mathcal{R}_q^o }{ 5 } \;
%\left[  
(1 + R_L)^{ -\frac{3}{4} } \; e^{ - \left( 
\frac{q \, d_D }{ a_L } \right)^2 } \mathcal{S} \left( \frac{q \, d_T}{a_L} 
\right)	
\; \sin \left[ { \frac{q \, d_H}{a_L} + \Delta \left( \frac{q \, d_T}{a_L} \right) }  \right] 
%\right] \;\; ,
\label{eq:F2_sol}
\end{equation}
where $ a_L =  a(t_L) = 1/(1+z_L) $, $ z_L = 1090 $, $ d_T = 0.1331 $ Mpc, 
$ d_H = 0.1351 $ Mpc, $ d_D = 0.008130 $ Mpc, 
$ d_A = 12.99 $ Mpc, and 
$ R_L \equiv 3 \Omega_B (t_L) / 4 \, \Omega_\gamma (t_L) = 0.6234 $ (the latest set of suitable parameters are taken from Planck 2018 \cite{planck18}).  
It is noteworthy at this stage to point out again that all three transfer functions are completely determined by standard measured cosmological parameters, 
so that the only remaining ingredient to fully determine the $ C_l^{TT} $ coefficient is the 
initial (or primordial) spectrum $  \mathcal{R}_q^o $, where $q$ is the wavenumber, and ``o'' refers to outside the horizon.  Conventionally, $  \mathcal{R}_q^o $ is parameterized by an 
amplitude $ N $ and spectral index $ n_s $,
\begin{equation}
\mathcal{R}^o_q = 
N \, q^{-3/2} \;  {\left( \frac{q}{ q_\mathcal{R} } \right) }^{(n_s-1)/2} \;\;  .
\label{eq:Rq_param}
\end{equation}
Here the reference ``pivot scale'' is usually taken to be 
$ q_\mathcal{R} = 0.05 \, \rm{Mpc}^{-1} $ by convention.  
As a consequence, once the primary function $ \mathcal{R}_q^o $ is somehow determined, 
classical cosmology is then expected to fully determine the form of the $C_l^{TT}$ spectral coefficients.
It is therefore possible to write the $ C_l^{TT} $'s fully, and explicitly, 
in terms of the primary function $ \mathcal{R}_q^o $.  
After expanding the plane waves factor in a complete set of spherical harmonics 
and spherical Bessel functions, $C_l^{TT}$ from Eq.~(\ref{eq:ClTT_invert}) becomes
\begin{equation}
C_l^{TT} = 16 \pi^2 \, T_0^2 \int_{0}^{\infty}{q^2 \,dq} \;
\left( \mathcal{R}_k^0 \right)^2
\left[ \, 
j_l ( qr_L ) \widetilde{F_1} (q) + j_l^\prime ( qr_L ) \widetilde{F_2} (q) \, 
\right]^2 \;\;  ,
\label{eq:Cl_jl}
\end{equation}
where $r_L = r\left(t_L\right) $, and we have factored out the function
$ \mathcal{R}_q^o $ explicitly by defining 
$ F_1 (q) = (\mathcal{R}_q^o) \, \widetilde{F_1} (q) $ 
and 
$ F_2 (q) = (\mathcal{R}_q^o) \, \widetilde{F_2} (q) $.  

Now, recall that the matter power spectrum $P_m (k)$ is given by
\begin{equation}
P_m \left( k \right) = C_0 \left( \mathcal{R}_k^0 \right)^2 \, k^4 
\left[ \mathcal{T} ( \kappa ) \right]^2 \; ,
\label{eq:Pk_R}
\end{equation}
which tells us that we can obtain a direct relation between the matter power spectrum $ P_m (k) $ and the angular temperature coefficients $C_l^{TT}$,
\begin{equation}
C_l^{TT}  =  16 \pi^2 \; T_0^2 \int_{0}^{\infty} q^2 dq \,\, P_m(q) 
\left[ \, C_0 \, k^4 \, { \mathcal{T} ( \kappa ) }^2 \, \right]^{-1}
\left[ \,
j_l ( qr_L ) \widetilde{F_1} (q) + j_l^\prime ( qr_L ) \widetilde{F_2} (q) \, 
\right]^2 \, ,
\label{eq:Cl_Pk}
\end{equation}
where $ q $ and $ k $ are related by $ q=a_0 k $, and the scale factor ``today''
$ a_0 $ can be taken to be $ 1 $.  As a result, the predictions on $P_m (k)$ can be directly transformed into a prediction for $C_l^{TT}$.
Utilizing the same parameters in the numerical programs, the effects of with and without the RG running of Newton's constant (with IR regulation) on $C_l^{TT}$ can then be generated.  

% Fig. 4

 Fig. \ref{fig:PubPlot3_ClTT} shows the numerical result of $c_0 = 0$ (no running) $1.146$, and $8.02$ with the blue, green, and orange curve respectively, generated by ISitGR.  The observational CMB data from Planck (2018), as well as an (error-weighted) cubic fit for ence, is also shown.
 Noticing that the point $l=2$ is anomalously low, with large uncertainty due to cosmic variance, the error-weighted fit shown in this plot has not included the $l=2$ point. 
 
 From Fig. \ref{fig:PubPlot3_ClTT}, we see that the effects of a RG running of Newton's constant generally cause an upturn to the spectrum at low-$l$'s, starting at roughly $l=20$.  It can also be seen that the orange curve with a quantum amplitude [see Eq.~(\ref{eq:Grun2})] $c_0=8.02$ (or $ \xi = 5300 $ Mpc) creates a much more dramatic deviation, reaching a maximum of $140\%$ larger in value compared to the blue, classical ($c_0=0$) $\Lambda$CDM curve, while the green curve with $c_0=1.146$ (or roughly $ \xi \simeq 2.65 \times 5300 = 14000$ Mpc) has a milder deviation of $\approx 24\%$ from the classical result.  Again, neglecting the anomalous $l=2$ point, the green curve with $c_0=1.146$ is generally consistent with all observational data, arguably also with the desirable feature of marginally going through the error bars of $l=5$ and $l=6$.  On the other hand, the orange $c_0=8.02$ curve, while still lying within a few points' error margins, is less favorably supported by the data in this plot.  It is also seen that its deviations starts earlier at a higher value around $l\sim 22$, which causes it to miss a few more error bars in the low $l$ points.  As a result, the numerical results of this $TT$ plot shows that the green $c_0=1.146$ (or $\xi \approx 14000$ Mpc) curve is currently a more favorable parameter than the orange one.  Note that this is also consistent with the discussion and conclusion from the matter power spectrum $P_m (k)$ plot in Fig. \ref{fig:PubPlot2_Pk}.
 
We also investigated the results with all 3 programs.  However unlike $P_m (k)$, the three programs do not agree, despite being supplied with the same RG modified expression for Newton's constant.  
Fig. \ref{fig:PubPlot3_ClTT} displays the result from ISiTGR, which seems to be the most consistent for all plots.  
MGCAMB produces a much more dramatic upturn effect from the RG running at small $l$'s, roughly having its $c_0 = 1.146$ curve coinciding with ISiTGR's $c_0 = 8.02$ curve, and the MGCAMB $c_0=8.02$ curve even higher.  
On the other hand, MGCLASS predicts a much milder upturn, with {\it its} $c_0 = 8.02$ curve coinciding with ISiTGR's $c_0 = 1.146 $ curve.  
I.e. MGCAMB seem to predict an upturn around 7 times larger than ISiTGR, while MGCLASS seem to predict an upturn that is 7 times smaller than ISiTGR.  
Given the blackbox nature of such programs, it is unclear of the cause of this different given that all programs where supplied the same modification in Newton's $G$.  These programs, designed for modified gravity models, are known to be less well-tested compared to their base program (CAMB, CLASS), and it may not be surprising that two (or all) of them may be incorrect.  
One consistency is that all three programs predicts an upturn at low $l$'s, just to a different degree, roughly $\pm 1$ order of magnitude.  
Hence, it is at best that we can conclude from these available programs that the RG running of Newton's $G$ causes an upturn to roughly the order of magnitude presented in Fig. \ref{fig:PubPlot3_ClTT}.

Perhaps even more intriguing is the disagreement with a naive analytical analysis.  
From Eq.~(\ref{eq:Cl_Pk}), the first order estimate is that since $C_l^{TT}$ is the (weighted) integral of $P_m (k)$ over all k, a smaller $P_m(k)$ caused by an RG running (c.f. Fig. \ref{fig:PubPlot2_Pk}) should cause a smaller value of $C_l^{TT}$.  
In fact, if one assumes the transfer functions are not affected by the quantum corrections, the integral Eq.~(\ref{eq:Cl_Pk}) can be performed numerically (as done in \cite{hyu19}), since the classical interpolating formulas for the transfer functions are known, which does show a downturn, as naively expected, instead of an upturn.  
This work utilizes programs that in principle modifies the initial Friedmann and Boltzmann equations from the beginning, and includes any effects of the RG modified Newton's $G$ into the solutions, and thus in principle more trustworthy.  
But given the opaque nature of such programs, it remains further investigations through a more detailed study of the entangled initial set of coupled differential equations to fully understand the disagreements between the programs and the first-order analytical result, as well as the disagreement, and hence the reliability, within the numerical programs.

Nevertheless, given that these programs represents the most sophisticated tools currently, it is still constructive to look at their predictions of the quantum effects on other modes and variables of the CMB.  
For example, the theoretical predictions for the percentage deviations for $c_0=1.146$ curve with the classical curve is $\sim 37\%$ on $P_m(k)$ at its further available data point, while only $\sim 24\%$ on $C_l^{TT}$.  
This reveals the fact that the quantum effects maybe more significant in different physical variables.  
So by studying the predictions for different auto- and cross-correlations of different varaiables, and compare them to potentially independent data (e.g. ground-based measurements of E- and B-mode polarizations as oppose to space-based measurements of CMB temperature), new constraints and insight may be deduced.  
We will present the analysis and results of the other spectra of interest to cosmology in the remaining of this section.

\subsection{Temperature-E-mode Power Spectrum $C_l^{TE}$}
\label{subsec:TE}

The next few most popularly studied correlations on the CMB are the so-called $E$- and $B$-type polarization modes.  Here we will give a brief recap of the physics, and present the numerical results of the quantum corrections from a RG running Newton's constant, later compared with the observational data.

Reacll that observations of the CMB photons not only reveal their intensity (i.e. temperature) from various directions, but also the photons' polarizations, which can result from scattering on free electrons either at the time of recombination, or during the later period of reionization.  Measurements on polarizations then reveal extra information in constraining the parameters arising from a running of Newton's constant.

Following notations in \cite{wei08}, CMB photons distributions are fully described through a number density matrix $n^{ij} (\mathbf{x,p},t)$, or, more usefully, the dimensionless version of its perturbation $J_{ij} (\mathbf{x},\hat{p},t)$ (referred to as the dimensionless photon intensity perturbation matrix), related to $n^{ij}$ via
\begin{equation}
J_{ij} (\mathbf{x},\hat{p},t) 
\equiv 
\frac{1}{a^2 (t)} \frac{1}{\bar{\rho_\gamma}  (t)} 
\int^\infty_0 
4 \pi \, p^3 dp \;
\delta n^{ij} (\mathbf{x},p \hat{p},t)
\; .
\label{eq:def_Jij}
\end{equation}
In a line-of-sight direction $\hat{n}$, $J_{ij}$ can be parameterized via
\begin{equation}
J_{ij} (\mathbf{x},-\hat{n},t) 
= 
\frac{2}{T_0} 
\begin{pmatrix}
\Delta T(\hat{n}) + Q(\hat{n}) & U(\hat{n})-i V(\hat{n}) & 0 \\
U(\hat{n})+i V(\hat{n}) & \Delta T(\hat{n}) - Q(\hat{n}) & 0 \\
0 & 0 & 0 \\
\end{pmatrix}
\; ,
\label{eq:def_Jijmatrix}
\end{equation}
where $Q, U$, and $V$ are three real functions of direction (with units of temperature), known as the Stokes parameters, describing the photon's polarizations.  Notice that the photon temperature perturbations are given by the trace
\begin{equation}
\frac{\Delta T(\hat{n})} {T_0} 
= \quarter \;  J_{ii} (0,-\hat{n},t_0)
\; .
\label{eq:deltaT_Jij}
\end{equation}
It is these Stokes parameters that are measured in current observations of the CMB.  But since the scattering of light by non-relativistic electrons does not produce circular polarization, one expects that all CMB photons will be linearly polarized, so that $J_{ij}$ is real, and therefore $V$ = 0.  For further convenience of in comparing with observations of 2-point functions, which respect spherical symmetry, it is useful to expand the Stokes parameters $Q(\hat{n})$ and $U(\hat{n})$ seen in a direction $\hat{n}$ in a series of functions $\mathcal{Y}_l^m(\hat{n})$
\begin{equation}
Q(\hat{n}) + i U(\hat{n})
=
\sum^\infty_{l=2}
\sum^{l}_{m=-l}
a_{P,lm} \,
\mathcal{Y}_l^m(\hat{n})
\; ,
\label{eq:def_aP}
\end{equation}
\begin{equation}
\mathcal{Y}_l^m(\hat{n})
\equiv
2 \, \sqrt{ \frac{(l-2)!}{(l+2)!} } \, 
e_{+i} (\hat{n}) \, e_{+j} (\hat{n}) \,
\tilde{\nabla}_i \tilde{\nabla}_j \, Y^m_l (\hat{n})
\; ,
\label{eq:def_scriptYlm}
\end{equation}
where the subscript ``$P$'' in the coefficient $a_{P,lm}$ stands for ``polarization'', $\mathbf{\tilde{\nabla}}$ is the angular part of the gradient operator, and $\mathbf{e}_{\pm} (\hat{n}) = (1,\pm i,0)/\sqrt{2}$ are the polarization vectors in the direction $\hat{n}$.  To further satisfy the reality condition, one defines the amplitudes
\begin{equation}
a_{E,lm}
\equiv
-\left( a_{P,lm} + a^*_{P,lm} \right)/2
\;\; 
, 
\;\;\;
a_{B,lm}
\equiv
i\left( a_{P,lm} - a^*_{P,lm} \right)/2
\; ,
\label{eq:def_aEaB}
\end{equation}
so that their correlation functions
\begin{equation}
\left \langle a^*_{T,lm} \, a_{T,l'm'} \right \rangle
= C_l^{TT} \, \delta_{ll'} \, \delta_{mm'}
\; ,
\label{eq:def_ClTT}
\end{equation}
\begin{equation}
\left \langle
a^*_{T,lm} \, a_{E,l'm'}
\right \rangle
=
C_l^{TE} \, \delta_{ll'} \delta_{mm'}
\; ,
\label{eq:def_ClTE}
\end{equation}
\begin{equation}
\left \langle
a^*_{E,lm} \, a_{E,l'm'}
\right \rangle
=
C_l^{EE} \, \delta_{ll'} \delta_{mm'}
\; ,
\label{eq:def_ClEE}
\end{equation}
\begin{equation}
\left \langle
a^*_{B,lm} \, a_{B,l'm'}
\right \rangle
=
C_l^{BB} \, \delta_{ll'} \delta_{mm'}
\; ,
\label{eq:def_ClBB}
\end{equation}
are real and rotationally invariant.  The above relations define the various angular power spectrum functions $C_l^{XX}$, where $X= T,E,B$.  
The superscripts $E$ and $B$ are referred to as $E$- and $B$-type polarization respectively, 
since spatial-parity inversion, $a_{E,lm} \mapsto (-1)^l \, a^*_{E,lm}$, 
and similarly for $a_{T,lm}$, whereas $a_{B,lm}  \mapsto -(-1)^l \, a^*_{B,lm}$.  
As a result of parity, there are no bilinear correlations between $B$ with 
either $E$ or $T$.  (i.e. $C_l^{TB}=C_l^{EB}=0$.)

% Fig. 5
\begin{figure}
\begin{center}
\includegraphics[width=0.90\textwidth]{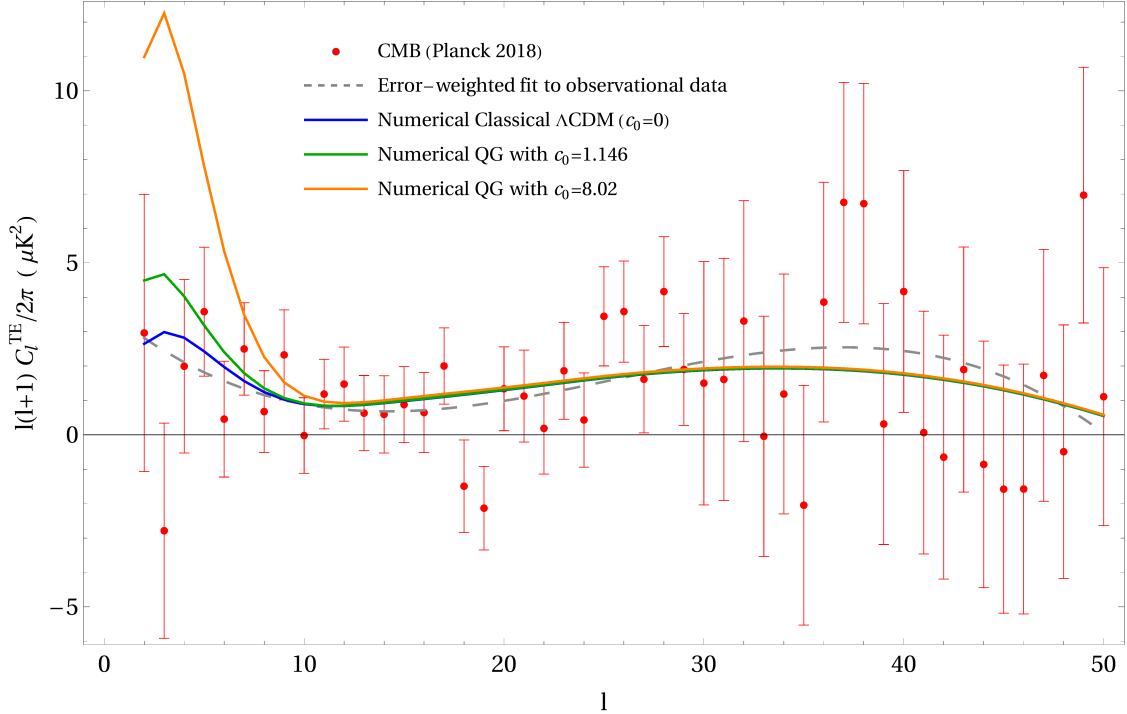}
\end{center}
\caption{
Comparison of the numerical prediction of the classical 
$\Lambda$CDM  program vs. the numerical predictions 
of the RG running of Newton's constant's effect on the cross-temperature-E-mode-polarization (TE) power 
spectrum $C_{l}^{TE}$.  
The solid curves represent the numerical predictions generated by the ISiTGR program, with the bottom (blue), 
middle (green), and top (orange) representing quantum amplitudes 
[see Eq.~(\ref{eq:Grun2})] quantum amplitudes [see Eq.~(\ref{eq:Grun2})] $c_0 = 0, 1.146, 8.02$ respectively. 
Here again one finds higher trends 
at large angular scales ($l<10$) as compared to the classical (no running) numerical 
$\Lambda$CDM  curve.  
The dashed curve represents an  error-weighted cubic fit to the observational CMB data from Planck (2018) \cite{planck18}.
}
\label{fig:PubPlot4_ClTE}
\end{figure}

With this background, we shall present the numerical predictions for the corresponding spectra with and without an RG running of Newton's $G$, compared against the latest observational data.
We start with the $TE$ spectrum.  Fig. \ref{fig:PubPlot4_ClTE} shows the numerical results with the observational data for $C_l^{TE}$.  The lowest solid (blue) curve represents the classical ($c_0=0$) spectrum, while the solid middle (green) and top (orange) curve represents the effect of a RG running Newton's constant with $c_0=1.146$ and $8.02$ respectively.  

It turns out new constraints for the RG running parameter $c_0$ can be deduced with this new plot.  With the inclusion of the $E$-type polarization data, we see that this has further constraints on some of the error bars in the low-$l$ data points.  This is due to the smaller error bars from the observational data in the $E$-type polarization correlations in the low-$l$ regime (see Fig. \ref{fig:PubPlot5_ClEE}).  As a result, one sees that the top $c_0=8.02$ curve (orange) is strongly disfavored by this plot.  
Another observation is that the difference between the $c_0=1.146$ and the classical $\Lambda$CDM ($c_0=0$) curve is about $60 \%$ in this $TE$ plot, which is a larger percentage deviation compared to $\approx 24 \%$ for the $TT$ plot.

We also compared the results from the other two programs (MGCLASS and MGCAMB, not shown on Fig. \ref{fig:PubPlot4_ClTE}). All the resultant curves of MGCLASS agree with ISiTGR for $l>30$, but for $l<30$, the $c_0=8.02$ curve of MGCLASS is about $36 \%$ lower than the corresponding ISiTGR curve.  All two curves with RG running from MGCLASS are within the error bars but due to the mismatch as shown in the Fig.\ref{fig:PubPlot1_P(m)compare}, MGCLASS results should be investigated further.  For MGCAMB, the curves with RG running are significantly higher than ISiTGR, making them disfavored. Also there is a slight horizontal shift for MGCAMB in $l$-space compared to the other two program, which should be investigated further.  
\subsection{EE- Power Spectrum $C_l^{EE}$}
\label{subsec:EE}

% Fig. 5
\begin{figure}
\begin{center}
\includegraphics[width=0.90\textwidth]{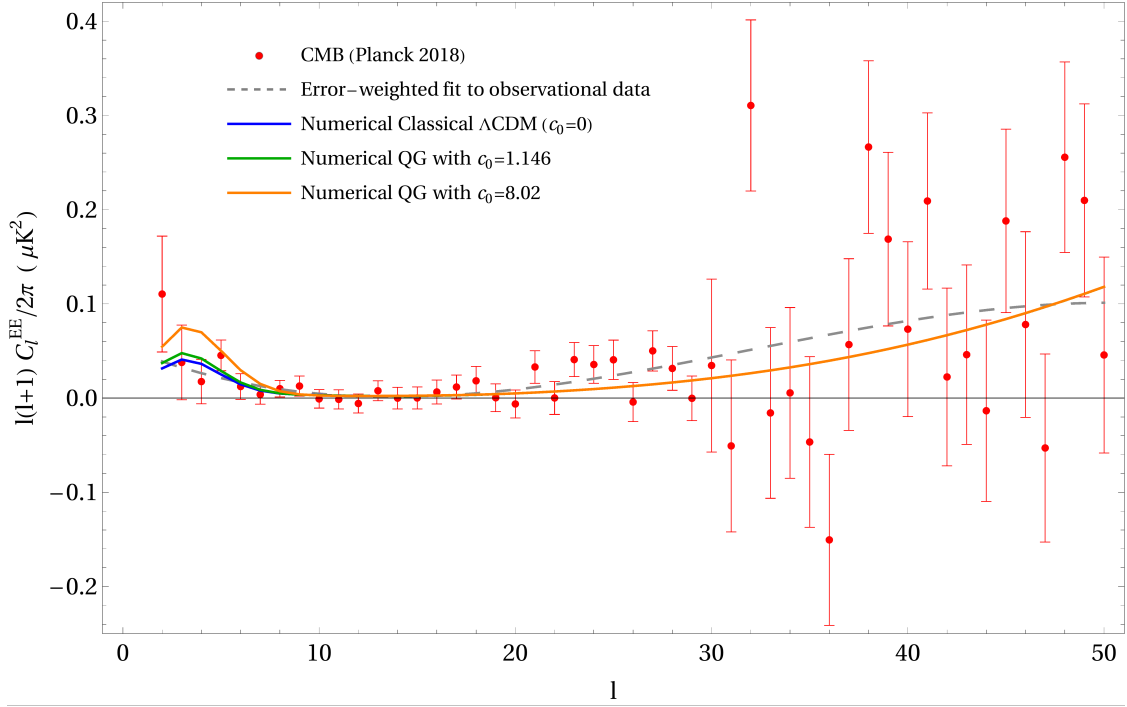}
\end{center}
\caption{
Comparison of the numerical prediction of the classical $\Lambda$CDM  program vs. the numerical predictions 
of the RG running of Newton's constant's effect on the E-mode (EE) power spectrum $C_{l}^{EE}$.  The solid curves 
represent the numerical predictions generated by ISiTGR, with the bottom (blue), middle (green), and top 
(orange) representing quantum amplitudes quantum amplitudes [see Eq.~(\ref{eq:Grun2})] $c_0 = 0, 1.146, 8.02$ respectively, showing slightly higher trends for large angular scales 
($l<10$) as compared to the classical (no running) numerical $\Lambda$CDM  curve.  
The dashed curve represents error-weighted cubic fit for observational CMB data from the Planck (2018) 
collaboration \cite{planck18}.
}
\label{fig:PubPlot5_ClEE}
\end{figure}

We move on to the $EE$ spectrum. Fig. \ref{fig:PubPlot5_ClEE} shows the numerical results with the observational data for $C_l^{EE}$.  
We also plotted an error-weighted cubic fit (dashed line) for the classical ($c_0=0$) spectrum (solid blue), as well as the quantum RG running of $G$ for the above values for $c_0$ (green and orange). It can be seen that there is no significant deviation from standard $\Lambda$CDM prediction like in temperature power spectra and all the curves are well within the data point error bars.
We can see that in the large scales ($l<20$) the errors are significantly small which makes T-E spectrum having smaller error bars in the scale of interest in this paper. 
% Weinberg pg379 eqn 7.4.40

When the other two programs are compared, there is no significant deviation to rule out any any curve. There is no noticeable deviation for MGCAMB curves from ISiTGR  for $l<20$ but there is a slight upward deviation for $l>20$. With MGCLASS, the RG curves are smaller than ISiTGR making smaller deviation from $\Lambda$CDM curve. 

%- Observation: smaller error bars in low-$l$ data.

%- Which contributes to making smaller error bars in the cross $TE$ spectrum

%- Which is what allows us to use it in Fig. \ref{fig:PubPlot4_ClTE} to disfavor the $c_0=8.02$ option. \\

%- Comment:  However, this plot on its own fails to rule out the $c_0=8.02$ curve, 

%- since there is very small distinction between within and without running.  \\

%- Discuss: Comparison with MGCAMB, MGCLASS.\\

\subsection{BB- Power Spectrum $C_l^{BB}$}
\label{subsec:BB}

\begin{figure}
\begin{center}
\includegraphics[width=0.90\textwidth]{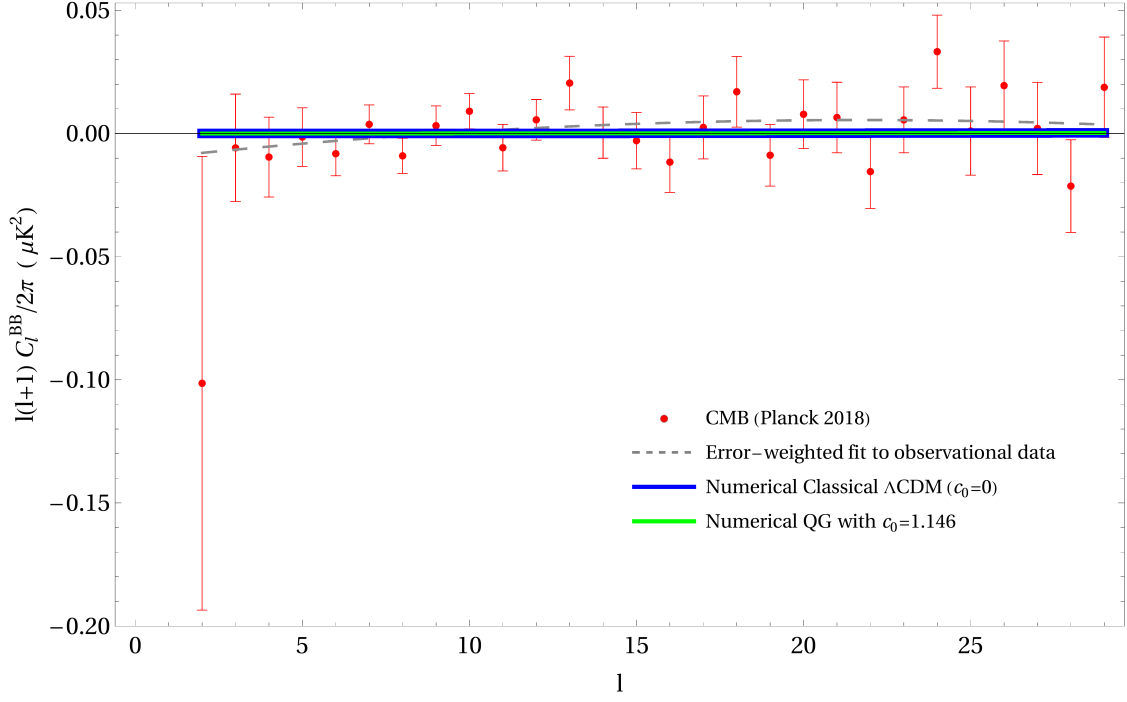}
\end{center}
\caption{
Comparison of the numerical prediction of the classical $\Lambda$CDM  program vs. the numerical prediction 
of the RG running of Newton's constant's effect on the B-mode (BB) power spectrum $C_{l}^{BB}$.  
The solid curves represent the numerical predictions generated by ISiTGR, with the top (blue) 
and bottom 
(green) representing $c_0 = 0, 1.146$ respectively, showing no significant deviation with the classical no running numerical $\Lambda$CDM  curve.  
The dashed curve represents error-weighted quadratic fit for observational CMB data from the Planck (2018) 
collaboration \cite{planck18}.  
The other $c_0 = 8.02$ curve, like the $c_0 = 8.02$ (green) curve, is consistent with zero (to around 1 
part in 10,000).  
So any deviations from the classical and $c_0 = 1.146$ curve are too insignificant to be seen, and negligible 
relative to the size of the error bars from the current latest data.  
Hence the $c_0 = 8.02$ curve is not included in this plot for clarity.
}
\label{fig:PubPlot6_ClBB}
\end{figure}

Next we discuss about B-mode polarization power spectrum, here shown in  Fig. \ref{fig:PubPlot6_ClBB}.
We have plotted an error-weighted quadratic fit (dashed line) for the classical ($c_0=0$) spectrum (solid blue), as well as the RG varying of Newton's $G$ for $c_0=1.146$ (green). It can be seen that there is no noticable deviation from standard $\Lambda$CDM prediction like in the temperature power spectra, and all the curves are well within the data point error bars. Because of the unnoticeable deviation, we didn't include the $c_0=8.02$ curve.
In standard cosmology, due to weak lensing there is a partial conversion of the E-mode to the B-mode polarization and it's predicted to be considerable around $l~1000$ scale which leaves large scale ($l<30$) close to zero. Due to limitations in dust modeling and telescopes limitations there is only data upto $l=29$. 

% Fig. 7

\subsection{Lensing Power Spectrum $C_l^{\phi\phi}$}
\label{subsec:PP}

The theory of CMB lensing is a vast topic on its own.  Here, we will try to present the key defining equations of the lensing spectrum, and then look at the numerical results of quantum gravitational effects on the lensing potential spectra.  A more complete account for the physics and observations can be found in \cite{wei08, eng14,planck18lens}. 

Consider a small deflection angle $\mathbf{\theta}$ from the undeflected direction $\hat{n}$ of a CMB photon, with $\mathbf{\theta}$ describing perpendicular direction to $\hat{n}$, and $|\mathbf{\theta} | \ll 1$.  Define the shear matrix $M_{ab}$ as
\begin{equation}
\Delta \theta_a 
=
\sum_b M_{ab} (r_S, \hat{n}) \, \theta_b
\; ,
\label{eq:Mab_def}
\end{equation}
where $a,b$ run over the directions orthogonal to $\hat{n}$, $r_S$ is the radial distance of the source from earth in a Robertson-Walker coordinate system, and $\delta\theta_{a}$ is the amount of deflection of $\theta$.  
From standard general relativistic calculations, the shear matrix can 
be related to the Newtonian potential of the lens source ($\phi$), via 
\begin{equation}
M_{ab} (r_S, \hat{n})
=
2 \, \int_0^{r_S} dr \, \frac{ r(r_S, \hat{n}) \, r }{ r_S }
\left[
\frac{ \partial^2 }{ \partial y_a \partial y_b }
\, \delta \phi ( r \hat{n} + \mathbf{y}, t )
\right]_{\mathbf{y}=0,t=t_r}
\;\; ,
\label{eq:Mab_expand}
\end{equation}
where $\mathbf{y}$ is a small perpendicular deflection vector to $\hat{n}$, and $t_r$ is the time for a photon that that just reached us from radial coordinate $r$.  Hence, the measurements of the shear matrix can yield information about perturbations to the gravitational potential ($\delta\phi$) by masses spread along the line of sight.
Define the so-called lensing convergence field $\kappa$ as
\begin{equation}
\kappa 
\, \equiv 
\frac{1}{2} \, \mathrm{Tr} M
\, =
\int_0^{r_S} dr \, \frac{ r(r_S, \hat{n}) \, r }{ r_S }
\left[
\left(
\nabla^2 - \frac{ \partial^2 }{ \partial r^2 }
\right)
\delta \phi ( r \hat{n} + \mathbf{y}, t )
\right]_{\mathbf{y}=0,t=t_r}  \;\; .
\label{eq:kappa_def}
\end{equation}
$\kappa$ is particularly useful because, if the lensing is due to a collection of bodies all at about the same radial coordinate $r_L$, it can be directly related to the matter perturbations $\delta \rho_m$.
More explicitly, $\delta \phi$ falls off rapidly for large distances, so that the factor $r(r,r_S) r$ can be replaced in a first approximation with $r(r_L,r_S)r_L$, and similarly the second term with ${ \partial^2 }/{ \partial r^2 }$ can be dropped.  Then Poisson's equation $a^{-2} \nabla^2 \delta \phi = 4 \pi G \, \delta \rho_m $ gives
\begin{equation}
\kappa =
\frac{ 4 \pi \, G \, a^2(t_{r_{L}}) \, d_A (LS) \, d_A (EL) }{d_A (ES)}
\int_0^{r_S} dr \;
\delta \rho_m ( r\hat{n} , t_L ) \, a(t_L)
\;\; ,
\label{eq:kappa_rhom}
\end{equation} 
resulting in an expression directly linking $\kappa$ to matter density fluctuations $\delta \rho_m$.  Hence, a measurement of the value of $\kappa$ for sources seen in one direction can reveal the total mass density of a cluster of lensing masses that lies along that line of sight at distance $r_L$ (projected onto a plane perpendicular to the line of sight).  
Since, as we have shown, gravity constraints the scaling of correlations of matter, it should also do so for $\kappa$.

So, to project the convergence field $\kappa$ onto the sky, we decompose it in a way that is analogous to the other angular spectra,
\begin{equation}
\kappa (\hat{n})
=
\sum_{lm} a_{\kappa,lm} \, Y_l^m (\hat{n})
\; ,
\label{eq:kappa_Ylm}
\end{equation}
with 
\begin{equation}
a_{\kappa,lm}
=
- 2 \pi \, i^l \int d^3 {\bf q} \, q^2 \; 
\alpha(\mathbf{q}) \, Y_l^{m*} (\hat{q})
\,
\int_0^\infty dr \,
g(r) \, \delta \phi_q (t_r) \left[ j_l (qr) + j''_l (qr) \right] 
\; ,
\label{eq:aklm_expand}
\end{equation}
with quantum noise fluctuation correlation 
\begin{equation}
\left \langle
\alpha(\mathbf{q}) \, \alpha^* (\mathbf{q'})
\right \rangle
=
\delta^3 (\mathbf{q}-\mathbf{q'})
\; ,
\label{eq:noise}
\end{equation}
which defines $C_l^{\kappa \kappa}$
\begin{equation}
\left \langle
a_{\kappa,lm} \, a^*_{\kappa,l'm'}
\right \rangle
=
\delta_{ll'} \, \delta_{mm'} \; C_l^{\kappa \kappa}
\; \; .
\label{eq:def_ClKK}
\end{equation}
Or more explicitly, by inverting the expression in Eq. ~(\ref{eq:def_ClKK}), 
\begin{equation}
C_l^{\kappa \kappa}
\, = \,
4 \pi^2 \,
\int_0^\infty q^6 \, dq \,\,
\left| \,
\int_0^\infty dr \, g(r) \, \delta \phi_q (t_r) \left[ j_l (qr) + j''_l (qr) \right]  
\, \right|^2
\; .
\label{eq:ClKK_invert}
\end{equation}
In the literature \cite{eng14, planck18lens}, it is often the correlation for the lensing potential $C_l^{\phi \phi}$ that is plotted, instead of that of the lensing convergence field $C_l^{\kappa \kappa}$, related by 
\begin{equation}
\kappa \, (\mathbf{\hat{n}})
\, = \,
\half \, \nabla^2 \phi (\mathbf{\hat{n}})
\;\; .
\label{eq:kappa_phi}
\end{equation}
And finally, the cross-correlations $C_l^{T \phi}$ and $C_l^{E \phi}$ can be similarly defined in analogous to Eq.~(\ref{eq:def_ClKK}) with respective expansion 
coefficients $a_{T,lm}$ and $a_{E,lm}$, similar to Eq. ~(\ref{eq:def_ClTT})-(\ref{eq:def_ClBB}).  
With this background, we will present the numerical results of including a 
quantum RG running of Newton's constant for these spectra.
%
% Fig. 8
\begin{figure}
\begin{center}
\includegraphics[width=0.90\textwidth]{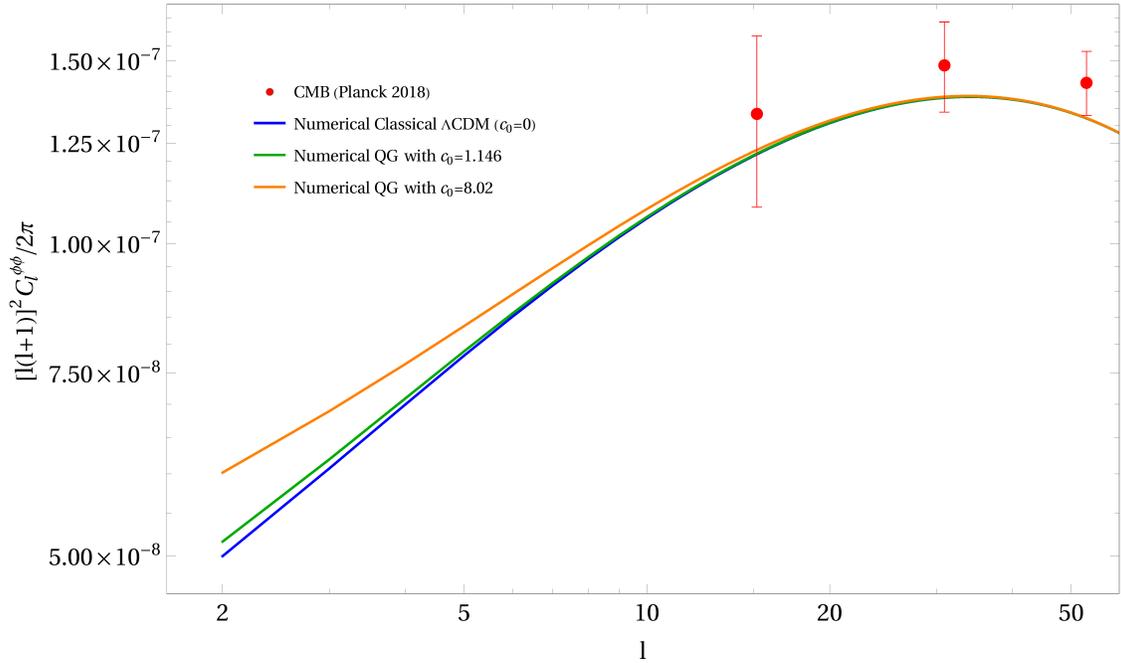}
\end{center}
\caption{
Comparison of the numerical prediction of the classical $\Lambda$CDM  program vs. the numerical predictions 
of the RG running of Newton's constant's effect on the deflection lensing ($\phi\phi$) power spectrum $C_{l}^{\phi \phi}$.  
The solid curves represent the numerical predictions generated by ISiTGR, with the bottom (blue), middle (green), 
and top (orange) representing the quantum amplitudes [see Eq.~(\ref{eq:Grun2})] $c_0 = 0, 1.146, 8.02$ respectively, showing just slightly higher trends at 
the very large angular scales ($l<5$) as compared to the classical (no quantum running) numerical $\Lambda$CDM  curve.
Only limited observational data is available currently, especially at large angular scales (below $l<15$).
}
\label{fig:PubPlot7_ClPhiPhi}
\end{figure}
%

%Cl_PP description

%complete the sentence
In Fig.\ref{fig:PubPlot7_ClPhiPhi} for $C_l^{\phi \phi}$ 
we have plotted for the classical ($c_0=0$) spectrum (solid blue), with as well as for an RG running of $G$ with the previously used values for $c_0$ (green and orange). One can see that there are a significant deviation which is up to $80\%$ for $c_0=8.02$, but only $10\%$ for $c_0=1.146$, compared to the standard $\Lambda$CDM prediction.
Due to current observational limitations there are only three data points which lie inside our region of interest ($l<50$). Apart from Planck collaboration (2018) data other projects such as the South Pole Telescope (SPT) \cite{Sim18} and the Atacama Cosmology Telescope (ACT) \cite{das14} have few observational data points which mostly lie in the region $l>100$. 

\subsection{Temperature-Lensing Power Spectrum $C_l^{T\phi}$}
\label{subsec:TP}

\begin{figure}
\begin{center}
\includegraphics[width=0.90\textwidth]{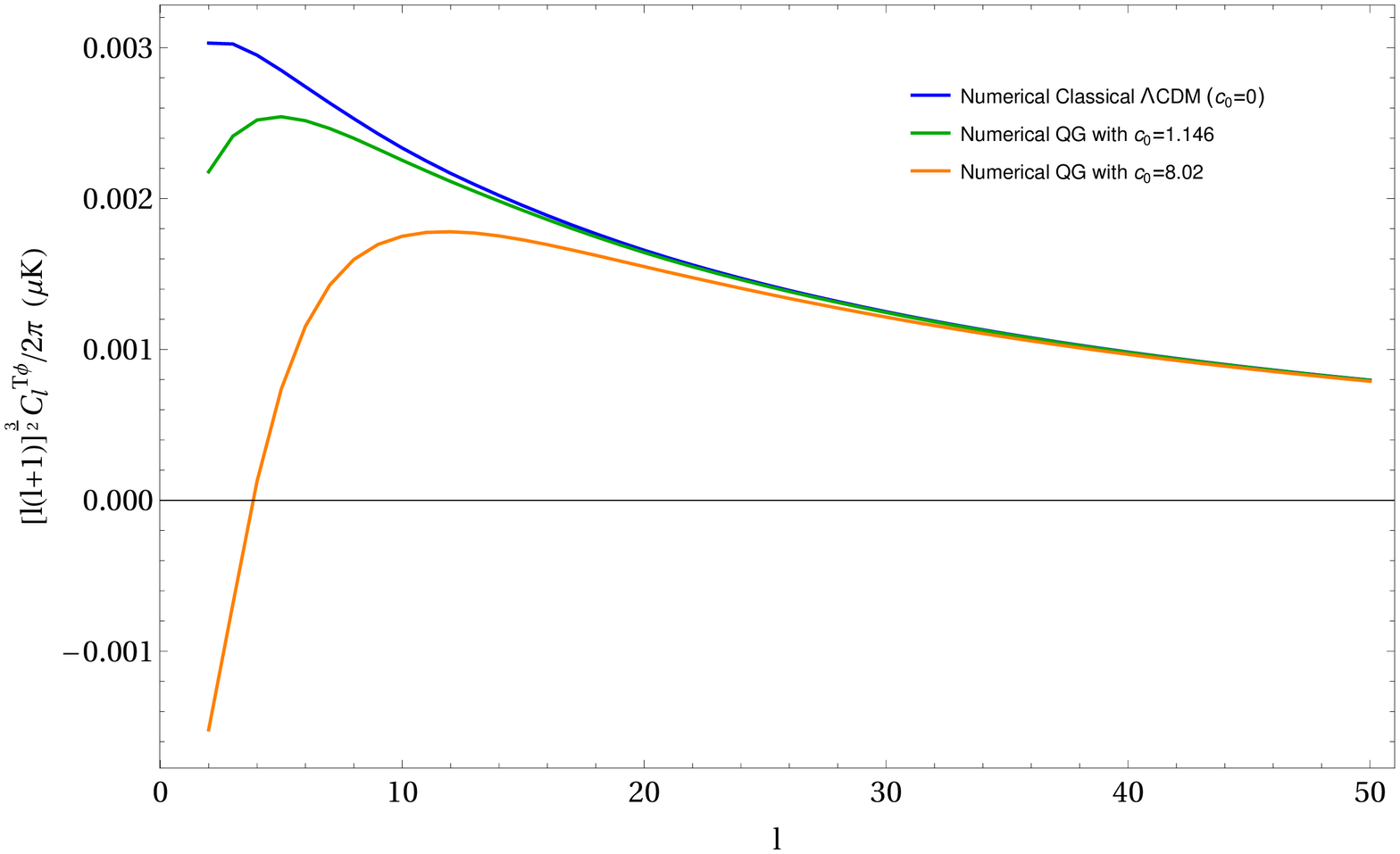}
\end{center}
\caption{
Comparison of the numerical prediction of the classical $\Lambda$CDM  program vs. numerical predictions 
of the RG running of Newton's constant's effect on the cross-temperature-lensing (T$\phi$) power spectrum 
$C_{l}^{T \phi}$.  
The solid curves represent the numerical predictions generated by ISiTGR, with the bottom (blue), middle (green), 
and top (orange) representing quantum amplitudes [see Eq.~(\ref{eq:Grun2})] $c_0 = 0, 1.146, 8.02$ respectively, showing again just slightly higher trends 
at large angular scales ($l<20$) as compared to the classical no running numerical $\Lambda$CDM  curve.  
No data with reasonable errors are found so far for $C_{l}^{T \phi}$.
}
\label{fig:PubPlot8_ClTPhi}
\end{figure}

For the $C_l^{T \phi}$ and $C_l^{E \phi}$ power spectra there are no observational data points so far, and given $C_l^{\phi \phi}$ having limited number of data points we don't expect to have any in the large scale region ($l<50$). 
In Fig. \ref{fig:PubPlot8_ClTPhi} we show $C_l^{T \phi}$ and
we have plotted the classical ($c_0=0$) spectrum (solid blue) with RG with the above values for $c_0$ (green and orange). It can be seen that there is significant deviation for $c_0=8.02$ which drops to negative values. The deviation begins for scales corresponding to $l<30$ . For $c_0=1.146$ it drops up to $30\%$ of standard $\Lambda$CDM prediction. 

% Fig. 8

\subsection{Lensing-E-mode Power Spectrum $C_l^{E\phi}$}
\label{subsec:EP}

\begin{figure}
\begin{center}
\includegraphics[width=0.90\textwidth]{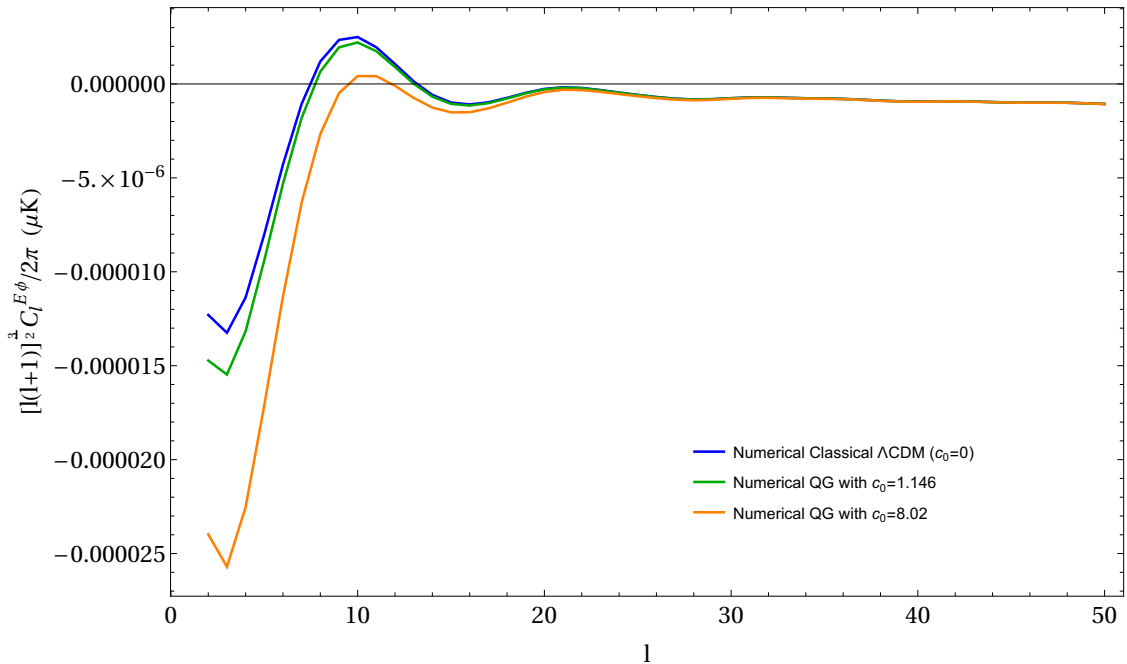}
\end{center}
\caption{
Comparison of the numerical prediction of the classical $\Lambda$CDM  program vs. numerical predictions 
of the RG running of Newton's constant's effect on the cross-E-mode-lensing (E$\phi$) power spectrum 
$C_{l}^{E \phi}$.  
The solid curves represent the numerical predictions generated by ISiTGR, with the top (blue), middle (green), 
and bottom (orange) representing quantum amplitudes [see Eq.~(\ref{eq:Grun2}) quantum amplitudes [see Eq.~(\ref{eq:Grun2})] $c_0 = 0, 1.146, 8.02$ respectively. showing this time smaller trends at the large angular scales ($l<20$) as compared to the classical no running numerical $\Lambda$CDM  curve.
No data with reasonable errors are found so far for $C_{l}^{E \phi}$.
}
\label{fig:PubPlot9_ClEPhi}
\end{figure}

In Fig. \ref{fig:PubPlot9_ClEPhi} we show the results for $C_l^{E \phi}$, and
we have plotted the classical $\Lambda$CDM ($c_0=0$)  spectrum (solid blue) compared with the RG running of Newton's $G$ spectrum with the above values for $c_0$ (green and orange). It can be seen that there is significant deviation for $c_0=8.02$ which drops by more than $50\%$ which gets ruled out. The deviation begins scales corresponding to $l<30$ . For $c_0=1.146$ it drops within $20\%$ of the standard $\Lambda$CDM prediction. Since for now there are limited observational data points for $C_l^{\phi \phi}$ nothing can be done about ruling out any specific model. But in the near future with CMB-S4 (The next generation "Stage-4" ground-based CMB experiment) \cite{cmbs4} more data on $C_l^{\phi \phi}$ , $C_l^{E \phi}$ might provide a good test for the models.

%CMBS-4 reference 

% Fig. 9

\vspace{30pt}

%%%%%%%%%%%%%%%%%%%%%%%%%%%%%%%%%%%%%%%%%%%%%%%%%%%%%%%%%%%%%%%%%%%%%
\section*{Conclusion}
\label{sec:conc}  

In this paper, we have revisited the derivation of the matter and temperature power spectra from the quantum theory of gravity without invoking any additional scalar fields from inflation, which, to our knowledge, is the first of its kind.  We reviewed that while the short-distance quantum theory of gravity remains speculative, the long-distance behaviors are well known and primarily governed by the renormalization group (RG) behaviors near its critical point.  In particular, we reviewed how the critical scaling dimension ``$s$'' of the correlation function of the scalar curvature fluctuations at large distances directly governs the scalar spectral index ``$n_s$'' of the cosmological spectra, as well as the additional quantum gravitational effects, such as the (IR-regulated) renormalization group running of the coupling constant (Newton's constant) $G$, that will affect these spectra subtly at large distances.  
We then presented the various numerical programs that we used in this work, and their main results, to complement the previous mainly analytical analysis.  We then utilized these programs to further study other cosmological spectra of different modes.  We compared these with latest available observational data, and provided new constraints and insights to the parameters ($c_0, \, \xi$) of the quantum theory.  
We also discussed the possibility of verifying, or falsifying, some of these hypothesis with increasingly powerful observational cosmology experiments in the future.

Using the numerical results, we find that especially the plots of the matter power spectrum $P_m(k)$, the angular temperature spectrum $C_l^{TT}$, and the angular temperature-E-mode spectrum $ C_l^{TE}$- all play an important role in revealing new insight to constraining the quantum amplitude $c_0$, a parameter that governs the size of quantum corrections due to the RG running of Newton's constant.  We find that all three plots agreeably favors a value of $c_0$ closer to around $1.15$, rather than the naive estimate of $\sim 8.0$.  This is in particular obvious in the new $C_l^{TE}$ plot from this work, with the $c_0=1.146$ curve showing a $\sim 60\%$ deviation from the classical $\Lambda$CDM (no quantum running) curve.  
On the other hand, the angular E-mode spectrum $C_l^{EE}$ and angular angular B-mode spectrum $C_l^{BB}$ plots are the least useful in distinguishing the running effect, with the $EE$ plot showing only a mild deviation of about 15\% from the classical prediction for the $c_0=1.146$ curve, and the deviations on the $BB$ plot are basically consistent with zero.  
The three angular lensing spectra, $C_l^{\phi \phi}, C_l^{T \phi}$ and $C_l^{E \phi}$, are potentially feasible candidates in providing further insights and constraints.  Especially for the $T \phi$ plot, showing around 20\% and almost 150\% deviation for the $c_0=1.146$ and $c_0=8.02$ curve respectively, from the classical curve.  However, all these latter spectra suffer from a lack of observational data in the low-$l$ regime, making it impossible to draw any conclusion about the favorability of the parameter or the RG running in general at this stage.  

However, although the percentages differences between the spectra with and without quantum corrections are decently significant for scales below $l<10$ -- ranging from $\sim 15-60 \%$ even with the milder value of $1.146$ for $c_0$, the uncertainties from current observational data in those ranges are unfortunately even larger.  As a result, it is not yet possible to conclude at this stage the visibility of these effects.  At best, one can claim the slight hints of RG running from the smallest data point in $P_m(k)$, as well as the last few points ($3<l<7$, ignoring the anomalous $l=2$ point) of $C_l^{TT}$.  
Nevertheless, with technology and precision of cosmological experiments improving at a rapid pace, better observational data in this regime perhaps forms one of the most promising area where quantum effects of gravity can be revealed and tested for the first time.  This is a consequence of the concrete predictions of the long-distance quantum effects, based on well-established renormalization group analysis, as opposed to the still rather speculative short-distance theories of gravity.

From a theoretical perspective, the numerical results from this work also serve an important purpose in ruling out the less favorable value of $c_0=8.0$ for the quantum amplitude, but instead suggesting a value around seven times smaller, closer to $c_0=1.15$.  We also noted that the uncertainties in the observational data at low-$l$'s cannot yet fully constrain the precise shape of the RG running, allowing for the possibility that these various deviations can all be mimicked instead by a modified value of $\xi \approx 14000$ Mpc, or around 2.5 times larger than the naive estimate $\xi \simeq \sqrt{3/\lambda} = 5300 \text{ Mpc}$.  
As we discussed in the theory section (Sec. \ref{sec:background}), unlike the universal critical scaling index $\nu$ (shown from various method to have a value very closed to $\nu = 1/3$), the parameters $c_0$ and $\xi$ do not necessarily follow from universality, but are instead confident only up to order of magnitudes.  
While the observational data at this stage cannot yet exhibit the effects of RG running, they do provide a useful constraint to the possible values of these theoretical parameters.  
In particular, as studied in detail in our earlier work \cite{hyu19}, even with the current observational data's precision, they provide an extremely stringent constraint on the allowed values of $\nu$, down to at most a $1-2\%$ deviation from $1/3$.  This result not only provides a great verification of the values obtained from various theoretical methods such as the Regge lattice calculations of the path integral, but perhaps the first phenomenological test of the quantum theory of gravity in cosmology.  
It is thus hopeful that as observational technology continues to improve, more insights can be gained regarding the values for $c_0$ and $\xi$.  
With more data and smaller error bars, one can further narrow down a best fit value for the quantum amplitude $c_0$ or vacuum condensate scale $\xi$ by Markov Chain Monte Carlo (MCMC) sampling in the ISiTGR program.  
In addition, ISiTGR is also capable of calculating tensor perturbations, which can be used to test this quantum gravitational picture as soon as more observational data on that becomes available.  
As a fundamentally tensor theory, this gravitational fluctuation picture is expected to produce nontrivial predictions to those of scalar field based inflation models.

It should also be noted that the numerical programs show a very encouraging agreement with the analytical results on the matter power spectrum $P_m(k)$, as shown here in Fig. \ref{fig:PubPlot2_Pk}.  This agreement provides great confidence in the analytical methodology used in \cite{hyu18}, or as summarized here in Sec. \ref{sec:background}.  
The concordance between numerical and analytical results provides extra support on how the quantum fluctuations of the gravitational field are linked to the fluctuations of the matter density field.  
However, the numerical results for the effects of a RG running of $G$, suggesting an upturn at low $l$'s, seem to disagree with the analytical intuition that a lower $P_m(k)$ should give a lower $C_l^{TT}$, as suggested in Eq.~(\ref{eq:Cl_Pk}).  
Since the derivation of Eq.~(\ref{eq:Cl_Pk}) is purely classical and does not involve any quantum gravitation input, this suggests a lack of analytical understanding of the effects of a having a modified RG running Newton's constant on the Boltzmann equations, and thus their solutions of the form factors $F_1$ 
and $F_2$ [Eqs. ~(\ref{eq:F1_sol}),(\ref{eq:F2_sol})].  
It is unclear analytically from the coupled differential equations how the running of Newton's $G$ from Eq.~(\ref{eq:promoteGrun}) affects their solutions, making it difficult to translate the predictions on $P_m(k)$, which agrees with the numerical results, to $C_l^{TT}$.  
This is an area under active further theoretical investigations, and will be addressed in future work.  
Nevertheless, armed with the supposedly more comprehensive and reliable numerical programs, new insights should be gained regarding the various quantum effects of gravity on the different cosmological spectra.

In conclusion, we have presented in this paper a compelling alternative picture for the various observed cosmological spectra that is motivated by gravitational fluctuations.  
In this work, we provided updated and extended analysis utilizing numerical programs in cosmology, as well as new physical predictions that can potentially distinguish this perspective from that of standard scalar field inflation.  
While inflation still currently forms one of the more popular approach, its full acceptance has remained controversial \cite{teg05,ste14}.  While there exists a number of alternatives to the standard horizon and flatness problems \cite{ste12,hol02}, the ability to explain the various cosmological power spectra has long been one of the unique predictions from inflation-motivated models, and thus often considered as one of the ``major successes'' for inflation.  
It is thus significant that this work provides an alternative, which is in principle arguably more elegant as it only uses Einstein gravity and standard nonperturbative quantum field theory methods, without the usual burden of flexibilities of inflation.
Nevertheless, because of the limited precision of current observational data, it is not yet possible to clearly prove or disprove either idea.  
Still, the possibility of an alternative explanation without invoking the scalar fields is significant, as it suggests that the observed power spectra may not be a direct consequence nor a solid confirmation of inflation, as some literature may suggest.  
By exploring in more details the relationship between gravity and cosmological matter and radiation both analytically and numerically, together with the influx of new and increasingly accurate observational data, one can hope that this hypothesis can be subjected to further stringent tests in the future.

\newpage

%%%%%%%%%%%%%%%%%%%%%%%%%%%%%%%%%%%%%%%

\vfill

\end{document}